\documentclass[10pt,preprint]{aastex}

\shorttitle{Flares with Frequency-Dependent Time Lags}
\shortauthors{Sokolov, Marscher, \& McHardy}

\received{2004 March 29}
\begin{document}

\title{Synchrotron Self-Compton Model for Rapid Nonthermal Flares in Blazars with Frequency-Dependent Time Lags}

\author{Andrei Sokolov and Alan P. Marscher} 
\affil{Department of Astronomy, Boston University, 725 Commonwealth Avenue, Boston, MA 02215, USA}
\email{sokolov@bu.edu, marscher@bu.edu}

\and 
\author{Ian M. McHardy}
\affil{Department of Physics and Astronomy, University of Southampton, Southampton SO17 1BJ, UK}
\email{imh@astro.soton.ac.uk}


\begin{abstract}
We model rapid variability of multifrequency emission from blazars
occurring across the electromagnetic spectrum (from radio to $\gamma$-rays).
Lower energy emission is produced by the synchrotron mechanism,
whereas higher energy emission is due to inverse Compton scattering
of the synchrotron emission.
We take into account energy stratification established by
particle acceleration at shock fronts and energy losses due to synchrotron emission.
We also consider the effect of light travel delays for the synchrotron emission
that supplies the seed photons for inverse Compton scattering.
The production of a flare is caused by the collision between
a relativistic shock wave and a stationary feature in the jet (e.g., a Mach disk).
The collision leads to the formation of forward and reverse shocks,
which confine two contiguous emission regions resulting in complex profiles of simulated flares.
Simulations of multifrequency flares indicate that relative delays between 
the inverse Compton flares and their synchrotron counterparts are dominated by energy stratification
and geometry of the emitting regions, resulting in 
both negative and positive time delays depending on the frequency of observation.
Light travel effects of the seed photons 
may lead to a noticeable delay of the inverse Compton emission with respect to synchrotron variability
if the line of sight is almost perfectly aligned with the jet.
We apply the model to a flare in 3C~273 and derive
the properties of shocked plasma responsible for the flare.
We show that the pronounced negative time delay between the X-ray and IR light curves
(X-rays peak after the maximum in the synchrotron emission) can be accounted for 
if both forward and reverse shocks are considered.
\end{abstract}


\section{Introduction}
In modeling emission variability from blazars, one generally assumes \citep[e.g.,~][]{mar98} that
the spectrum is dominated by synchrotron emission at lower energies (radio through UV)
and by inverse Compton emission at higher energies (X-rays and $\gamma$-rays).
One popular model explains blazar emission as the product of electrons
accelerated to very high energies by relativistic shocks originating somewhere
near the base of a relativistic jet and then traveling downstream
\citep{blandford79,mar85,hughes85}.
Emission from the accretion disk, hot corona, and molecular torus may contribute to
the quiescent levels of observed radiation but should not affect the most rapidly variable component
of emission, which has a time scale as short as $\la1\,\mbox{day}$.
The flux and level of variability of the jet is greatly enhanced by
relativistic and light-travel effects 
if the angle between the jet and the line of sight is small, as expected in the case of blazars. 

Despite considerable observational and theoretical progress in understanding
emission variability in recent years, there is still little agreement about
a number of essential features of the problem.
One of these is an important aspect of the inverse Compton emission: several sources of
seed photons may contribute significantly
to the observed X-ray and $\gamma$-ray flux. 
Among the possible sources  are synchrotron photons produced locally 
or in the co-spatial regions (synchrotron self-Compton, or SSC) and
external emission (external radiation Compton, or ERC) 
that can come from a variety of sources such as a molecular torus \citep{bla00},
an accretion disk \citep{der93}, and emission-line clouds \citep{sik94}.
In addition, 
\citet{ghi96} suggested that
locally produced synchrotron emission may be scattered by nearby clouds surrounding
the jet and enter the excited region once again 
where they can be scattered by relativistic electrons (mirror Compton, or MC).

The location as well as the trigger of the variable high energy emission is currently an open question.
Propagating shock waves can produce long-term (months to years) outbursts \citep{mar85,hughes85}.
At high frequencies,
interaction with a turbulent ambient medium in the jet can generate
more rapid fluctuations \citep{mar85,tra92}.
The onset of a flare could also be triggered by the collision between a  relativistic shock wave
traveling down the jet and a more slowly moving \citep{spa01} or stationary compression, 
e.g., the core seen  in  very long baseline interferometric (VLBI) images of blazars.
Here we consider this last possibility,
that a moving shock collides with a stationary formation associated with the core of the jet,
which results in rapid variability on the time scale of several days.

In order to progress with such models, the nature of the radio core needs to be defined.
Synchrotron self-absorption determines the location of what appears to be the core at lower radio frequencies,
manifested by an upstream shift in position as one increases the frequency of observation.
However, at sufficiently high radio frequencies ($\ga40\,\mbox{GHz}$),
we expect this trend to cease so that  images reveal the ``true'' core that corresponds
to a stationary physical structure in the jet. 

Cross-correlation analysis of long term variability of such blazars as PKS 1510$-$089, 3C~273,
and 3C~279 reveals strong correlations between high (X-rays) and low (optical, IR, and radio) energy emission,
suggesting that both synchrotron and inverse Compton emission arise from the same
location in the jets \citep{mch99,mar04}.
Moreover, a comparison of multi-epoch VLBI images
and long term light curves of these blazars
suggests that the flares are associated with activity in or near the radio cores.
Bright flares usually correspond to an increase in brightness of the 
radio core followed by the appearance of a superluminal feature emerging and traveling downstream from the core.
This analysis supports the notion that the flares may be generated by a collision
between a moving shock wave and a stationary structure.

Theoretical considerations \citep{cou48} 
as well as a number of numerical simulations of relativistic jets \citep{dub93,gom97}
have shown that
stationary structures may appear in jets as
a result of interaction between the ambient medium and the jet
if there is local pressure imbalance between the two.
This interaction leads to the formation of a system of standing oblique shock waves,
perhaps terminating in a strong shock (Mach disk) that lies  perpendicular to the jet flow.
A relativistic flow of plasma crosses the Mack disk, which
accelerates the plasma electrons to highly relativistic energies.
Synchrotron radiation produced by these particles then gives rise to the emission 
seen as the radio core in VLBI images.

In this paper,
we present a theoretical model for rapid emission variability from blazars based on the scenarios sketched  above.
We then discuss the properties of the model that can help constrain the physical properties of the emission
regions.
We apply the model to a flare from the quasar 3C~273 observed simultaneously in the infrared K-band and in X-rays
with RXTE in March 1999.
We fit the light curves and the evolution of the X-ray spectral index
during the flare and determine the physical parameters of the medium producing the emission.

\section{Previous Observational and Theoretical Work}
\subsection{Multifrequency Light Curves and VLBI observations}
Since 1996, \citet{mar04} have been conducting
multifrequency monitoring of the quasars PKS~1510$-$089 and 3C~279
with RXTE in the 2.4--20~keV X-ray band  as well as in several other spectral bands,
including radio and optical. 
The quasar 3C~273, which exhibits some blazar characteristics, was also observed simultaneously
in the X-rays and the infrared K-band during concentrated campaigns in January 1997 \citep{mch99}
and March 1999.
Cross-correlation of the light curves shows that there is a correspondence 
in variability between different bands.
This is particularly pronounced in two flares in PKS~1510$-$089 observed in 1997.
In addition, violent activity of 3C~279 in 1998 and 2001 in X-rays is well matched by 
contemporaneous variations in the optical.
There is also a general correspondence between the ejection of superluminal components
from the core and the onset of the flares.

Stressing parallels between AGNs and microquasars in our galaxy, \citet{mar02}
conducted a study of long-term X-ray light curves and VLBI maps of 3C 120.
They determined a minimum distance of $0.3\,\mbox{pc}$ between the X-ray source---identified with the
accretion disk for this particular galaxy---and the core of the radio jet.
The radio core is likely to be even farther downstream for higher luminosity AGNs that
harbor more massive black holes in their nuclei.

These observations  suggest a model of blazar variability in which both synchrotron (optical, IR, radio) and
inverse Compton emission (X-rays) originate from the same region or contiguous regions,
which are located in the relativistic jet at a site near the radio core seen in VLBI images. 
It is natural to assume that the onset of a flare is triggered by the collision between a moving shock wave and
the stationary structure in the jet (e.g., Mach disk) responsible for the radio core.
The excited plasma resulting from the collision moves down the jet and is later seen
as a superluminal feature.
We develop such a model in Sections \ref{model.gas}--\ref{model.properties} below.

\subsection{Acceleration of particles by relativistic shock waves}
Radio to infrared flares in blazars are usually explained by appealing to the shock in jet model
\citep{blandford79,mar85, hughes85}, which postulates that relativistic shock waves
in the jet cause particle acceleration
and nonthermal emission, appearing as superluminal knots in VLBI maps.
Relativistic shocks in a jet may result from intermittent variability in the physical properties 
of the portion of the central
engine that is connected with the formation of the jet.
The details of the jet production and collimation have been associated with differential rotation 
in the inner accretion disk, which creates a magnetic coil that expels accreting material \citep{mei01}.
It is likely that the quasi-steady process that leads to jet production is 
sometimes punctuated by dramatic events due to instabilities occurring in the accretion
disk surrounding the black hole.
One of these events may momentarily saturate the jet with extremely energetic plasma with much
higher pressure than the steady jet plasma downstream.
The high energy plasma will sweep up slower moving material, forming a shell of enhanced density that is
separated from the steady jet in front of it by a shock wave.
This is usually referred to as a ``forward shock''
since the shock front advances ahead of the  shocked material.
On the upstream side, 
the shell is separated from the high energy plasma that pushes it by a contact discontinuity.
Inside the energetic plasma, the information about its interaction with the slower flow ahead
is transmitted by a compression wave that can steepen into
a reverse shock.
The exact outcome depends on the details of the jet production and the instability that
injects energetic plasma into the steady jet \citep{bic02,mar94}.

A Fermi-type acceleration mechanism is typically invoked to explain how particles
gain the high energies required by observations of the synchrotron and inverse Compton 
emission \citep[see, e.g.,][]{kro99}.
The effects of relativistic beaming restrict the conditions
under which the repeated cycles of acceleration may proceed, but the resulting
energy spectrum of the accelerated particles is similar to that observed in
the non-relativistic case.
\citet{ach01}  performed stochastic simulations of the acceleration 
process at ultrarelativistic shock fronts and found that a power law spectrum
$N(E)\propto E^{-s}$ is produced with a nearly universal value of the slope, $s\approx 2.2$--$2.3$.
The maximum energy attainable in the acceleration process is limited by the
particle diffusion that eventually puts it outside of the restricted range of conditions 
needed for the acceleration cycles to proceed.
Moreover, extremely high energy particles produced by acceleration should suffer from severe
radiative energy losses through synchrotron and sometimes inverse Compton emission.
Radiative losses can steepen the energy spectrum of relativistic particles 
above a certain break energy and also
limit the maximum value of the energy in the spectrum.

\subsection{Stationary shocks in Gas Dynamical Simulations of Jets}
A stationary shock wave system in a relativistic jet
can be caused by pressure imbalance between the jet and the surrounding ambient medium.
Oscillations of the jet width occur as the
overpressured medium in the jet expands until its pressure falls below the ambient gas
pressure.
This underpressured plasma then  contracts under the influence of the external medium until
high pressure is restored.
This sets up a stationary system of oblique shocks and a Mach disk
if the pressure mismatch is substantial \citep[see][]{cou48}.

\citet{dal88} used the method of characteristics in two dimensions to determine 
that standing shocks appear in relativistic jets in a similar fashion to those in non-relativistic flows.
They speculated that the radio core is produced by plasma passing through 
an oblique shock/Mach disk system in the jet.
In this scenario, stationary hot spots seen downstream of the core \citep{jor01} may be explained
by the oscillatory behavior of the jet producing multiple sets of
stationary structures.
\citet{dub93} also employed the method of characteristics in order to study
the steady-state structure of relativistic jets with toroidal magnetic field.
They found that for constant external pressure the jet has strong periodic
features with large pressure and velocity changes.
These authors noted that the features could be associated with a locally strong toroidal magnetic field.
In their simulations, the strength of the features downstream decreased substantially
when the radial pressure gradient of the external medium was introduced.
Using a numerical relativistic fluid dynamics code, \citet{kom97}
modeled the appearance of jets on parsec scales by including moving shocks 
and a stationary reconfinement shock.
They identified moving shocks in their simulations as the observed superluminal knots 
and the reconfinement shock as the brightest stationary knot, i.e.,\ the radio core.

\citet{gom97} conducted hydrodynamical simulations of the generation, evolution, and
radio emission of superluminal components in axially symmetric relativistic jets.
Their simulations showed that an increase in speed of the flow at the jet base
results in a relativistic shock wave propagating down the jet.
Prior to the introduction of velocity perturbations, their calculations produced a steady jet
that  showed a periodic system of stationary oblique shocks
when the jet was initially overpressured with respect to the external medium.
\citet{gom97} notice that when the moving perturbation passes through a standing shock
the latter is dragged downstream for some distance before returning to
its initial position as the steady jet is reestablished.
Three-dimensional hydrodynamic simulations of relativistic jets \citep{alo03}
show more complex behavior when precession of the jet is introduced.

According to the simulations by \citet{gom97}
the Mach disk is expected to have a small size compared to 
the cross-section of the jet. Most of the plasma in the jet interacts with 
the oblique shocks, which decrease the bulk Lorentz factor 
by a moderate amount, after which the flow reaccelerates.
Even the plasma that passes through the Mach disk radiates only a small 
fraction of its internal energy before it reaccelerates into the 
lower density region downstream and once again reaches relativistic speed.
It can do so  as long as the slope of the electron energy spectrum
is steeper than $2$ or if the protons carry most of the kinetic energy.
Otherwise, a large fraction of the total energy will be dissipated through
radiation by the highest energy electrons.
Therefore, the radio core in most sources is not expected to be nearly 
as intrinsically luminous as the hot spots 
in the external radio lobes where most of the relativistic energy of 
the jet is deposited. In some sources, notably TeV blazars, most of the energy 
in the jet indeed seems to be dissipated on parsec scales, which is 
an expected outcome of our model if the distribution of electrons is 
flat with slope $s<2$ and the protons do not carry the bulk of the total
energy out to parsec scales.

The association of the radio core with the underlying standing structure 
in the jet is complex. It cannot be equated with the plasma slowed down 
and compressed by the Mach disk since this is very weakly beamed. 
VLBI observations of single 
radio cores suggest that the emission from radio cores is beamed in 
the direction along the jets. The sources with jets in the plane of the sky
do not possess strong radio cores, which again suggests relativistic 
beaming.
The contributions from oblique shocks and 
also from the plasma shocked by the Mach disk after it reaccelerates to 
relativistic speeds must be taken into account.

\subsection{Progress in Modeling}
In this section we review recent progress in theoretical understanding of
multifrequency variability of blazars.
We pay special attention to the basic assumptions made in the simulations
about the geometry, acceleration, and internal structure of the emitting regions.

In simulating the spectra of $\gamma$-ray emitting blazars, \citet{mas95} \citep[see also][]{mas97}
assumed that the emission originates in a spherical, homogeneous region into which
relativistic electrons with a power law energy distribution are injected.
They approximated that the injection has a fixed form without
identifying a specific acceleration mechanism but interpreted the overall picture
as one in which a shock front rapidly accelerates the electrons.
The light-travel time of the photons across the emitting region was neglected.
Changes in maximum energy of electrons $E_{max}$, magnetic field $B$, or the normalization
factor in the distribution of electrons $N_{0}$, were considered by these authors 
as the factors that cause the flares.

\citet{der97} modeled flares by computing the time dependent spectral flux 
of a relativistic, homogeneous, spherical blob of plasma
with instantaneous injection of relativistic electrons.
The flares considered were restricted to ones in which the energy loss of relativistic electrons 
is dominated by synchrotron or external Compton emission.
The external radiation that drives the inverse Compton scattering in the ERC model was
assumed to be isotropic, homogeneous, and constant, similar to what is expected
when the emitting blob is inside the broad emission line region (BLR).
A $\delta$-function approximation for Compton scattering and synchrotron emission was used
in computing the observed fluxes. 
Energy losses due to adiabatic expansion were neglected in this work.
\citet{der98} presented the results of the modeling of multifrequency flares 
based on the approach presented in \citet{der97}, stressing the correlations between 
optical and X-ray emission.

\citet{cop99} presented sample SSC calculations based on a one-zone
homogeneous model developed by \citet{cop92} that mostly concentrated on an improved treatment
of the microphysics at the expense of more relevant but poorly understood dynamical and
structural considerations.
A delay in Compton flares observed in the simulations was explained by
Compton radiation at relatively low frequencies being emitted by lower energy 
electrons with  longer lifetimes compared to higher energy electrons that produce
high frequency synchrotron emission.
They also note that in the SSC model the target photon intensity in the source 
can only change on a time scale defined by the light travel time through the source
or longer.

\citet{chi99} pointed out that if the electron distribution changes on
a time scale shorter than the light crossing time through the emission region, then
even a homogeneous source, i.e.,\ one with constant density of relativistic electrons throughout, 
will resemble an inhomogeneous one.
They stressed that considering only the time evolution of the energy distribution 
of the emitting electrons is not enough for correctly modeling the flares.
Due to light travel effects, the observer will see radiation produced in
separate parts of the source at different times and, therefore, characterized
by particle distributions having various ages.
Both simultaneous injection of electrons through the entire source and 
energization occurring at a shock
front traveling through a section of the jet were considered.
The emitting region was divided into parallel slices to account for light travel effects.
In the case of the shock injection, each slice was further divided into separate cells
occupied by particles of different age.
The angle between the jet axis and the direction toward the observer
was assumed to be $\theta\approx1/\Gamma$, where $\Gamma$ is the bulk speed of the shocked plasma.
The authors noted that the inverse Compton emission calculation requires consideration of
light travel effects as well, since the seed photons in the SSC model may travel across the source
on the same time scale as the scale of variability.
This could lead to a time delay of the Compton emission with respect to the synchrotron.
They did not, however, include the effect in their calculations.
The authors also stressed that the symmetric shapes of the light curves strongly constrain
the injection and cooling time scales.

\citet{kat00} computed time dependent SSC models to study the spectral evolution
and variability patterns of blazars based on a one-zone homogeneous approximation
with electrons injected uniformly throughout the emission region.
They quantitatively compared the model with 
observed X-ray flares to derive the physical parameters of the emitting region.
However, only the X-ray bands were used because the source was not observed in other
spectral regions during the flare.
Similar to the work by \citet{chi99}, the authors included the effects
introduced by radiation travel times of photons originating in different parts of the emission region,
but neglected the light travel delays of the seed radiation for the inverse
Compton scattering.

An illustration of the importance of light travel delays
in estimating the observed variability of emission from blazars was demonstrated
in \citet{sal98}, who suggested a model for explaining
an extremely rapid burst observed from Mrk 421.
In their model, the observed radiation is produced as the result of 
prolonged interaction between a disturbance traveling down the jet and conical shocks. 
Due to the specific conical geometry of the shocks chosen by the authors, the photons
that are emitted in the observer's direction arrive at the observer at essentially
the same time, producing a rapid, extremely intense burst of radiation.

Some models \citep[see][]{mes02} interpret gamma-ray bursts (GRBs)
as shells of plasma ejected from a center site, colliding as they move outward.
A system of forward and reverse shocks is established as a result of the collision.
These shocks accelerate plasma particles, which in turn radiate their energy and
produce a multifrequency flare.
\citet{spa01}  applied this ``internal shock'' scenario to blazar variability.
They attributed the formation of shells to intermittent variability of the central engine.
Plasma characteristics were treated not as parameters but rather as the result of the jet dynamics.
Despite appealing to forward and reverse shock waves, the authors assumed that
the emitting zone is homogeneous with the same energy distribution throughout
the entire emitting volume.
Light travel delays across the emitting region were neglected in their study.

In modeling broad-band spectra of blazars that include SSC emission as well as
ERC from both the BLR and the dusty torus surrounding AGN, 
\citet{bla00} found that the BLR 
may be less significant as the source of the seed photons than the radiation from hot dust in the torus
if the emission region in the jet is located at a distance greater than $\sim0.1\,\mbox{pc}$ from the 
central engine.
Reverberation mapping techniques \citep{pet93} indicate
that the average radius of the BLRs  is $\sim0.1L_{42}^{1/2}\,\mbox{pc},$ 
where $L_{42}$ is UV luminosity normalized by $10^{42}\,\mbox{erg}\,\mbox{s}^{-1}\,\mbox{\AA}^{-1}$.
Therefore, if the emitting shell in the jet is outside the BLR, then
the external radiation from emission line clouds will be Doppler boosted by
a lesser amount compared to the boosting of emission from the torus. 
The results of the modeling \citep{bla00} showed that the ERC light
curve may still be dominated by the contribution from the BLR at the highest observed
$\gamma$-ray energies simply due to the higher frequencies of the BLR photons
compared with the IR photons from the dust.
Further modeling by \citet{sik01} gave strong support to the SSC mechanism 
as the primary emission process in the X-ray band.

\section{Gas Dynamics of Shock Collision}
\label{model.gas}
In our model, calculations of the production and transfer of the synchrotron and inverse Compton emission 
are performed in the rest frame of the emitting plasma.
The input parameters that describe the geometry of the source and the state of the plasma are defined 
in the plasma rest frame as well.
If the bulk speed of the emitting plasma is given, the observed quantities such as
synchrotron and inverse Compton flux as a function of frequency and time can be calculated
provided that
the angle between the jet axis and the line of sight toward the observer is known.
Since variability of blazars seems to be associated with the radio core, we assume that
the collision between a moving shock and a stationary shock is the primary source
of multifrequency flares and outbursts.
In this section, we describe how the properties of the shocked plasma relate
to the conditions in the quiescent jet flow and the colliding shocks.

\begin{figure}
\plotone{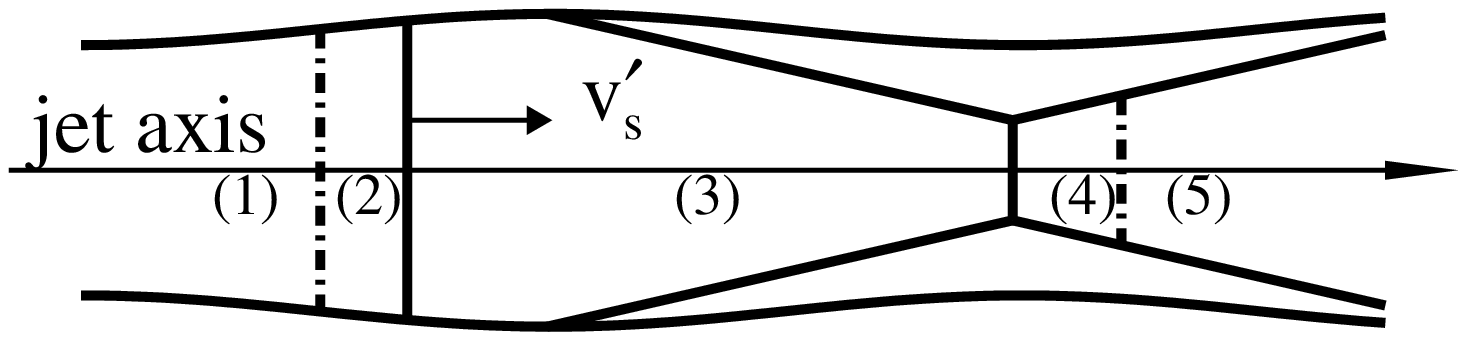}
\caption{A cross-sectional diagram of a jet with a shock wave moving to the right
and a stationary feature prior to collision. 
The moving shock is propelled by a fast stream of plasma
in zone (1). Zone (2) contains the shocked quiescent jet material piled up in front
of the fast flow. The shock between zones (2) and (3) moves at speed $v_s'$ in the AGN rest frame.
The Mach disk slows down the quiescent flow [zone (3)], creating a dense region downstream [zone (4)].
The plasma that has been slowed down by the Mach disk is eventually reaccelerated to relativistic
flow speeds in zone (5).
The transition between (4) and (5) is probably gradual, but a well-defined boundary is used
in the calculations for the sake of simplicity.
\label{jet}}
\end{figure}

\begin{figure}
\plotone{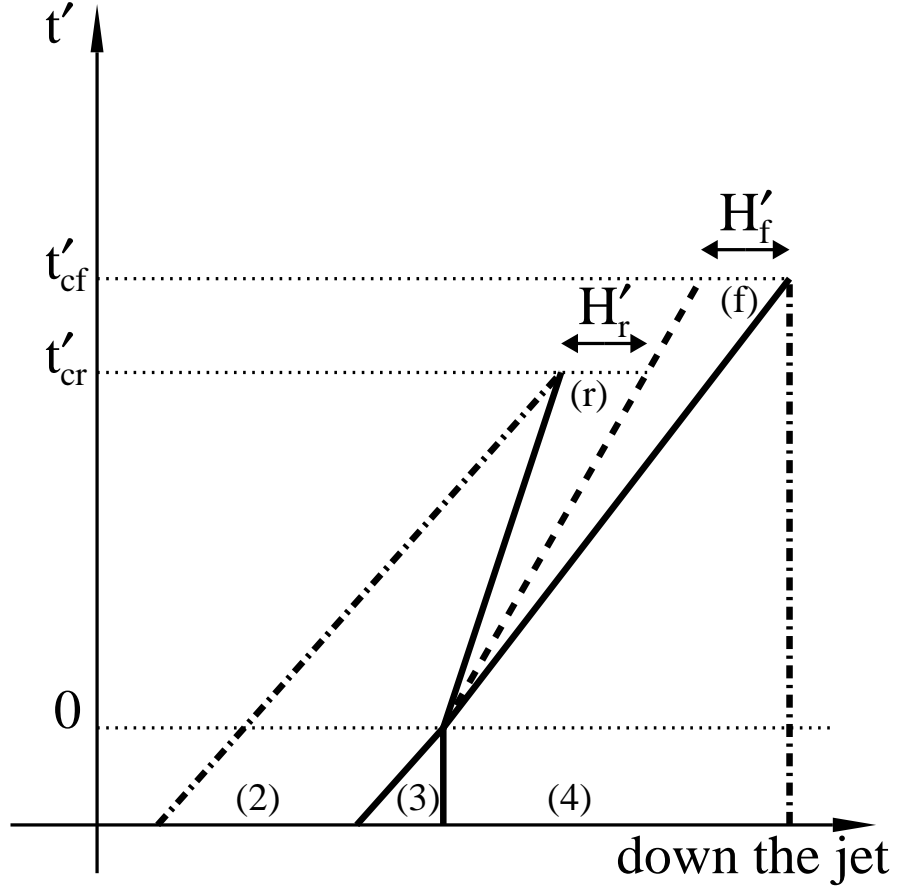}
\caption{The space-time diagram  of a collision between a stationary Mach disk and
a shock moving down the jet with Lorentz factor $\Gamma_s'=9$; 
the speed of the quiescent flow in zone (3) is $\Gamma_3'=4$.
Zones $(2)$, $(3)$, and $(4)$ are the same as in Fig.~\ref{jet}
(the positions of the shocks shown in Fig.~\ref{jet} corresponds to time before $t'=0$).
The collision occurs at time $t'=0$ at the location of the Mach disk.
As a result of the collision a system of forward and reverse shocks is set up (solid lines for $t'>0$).
Two new zones, $(r)$ and $(f)$, contain the material originally from zones $(2)$ and $(4)$
that has been shocked by the reverse and forward shocks, respectively.
Zones $(r)$ and $(f)$ are separated by a contact discontinuity (dashed line).
The size $H'$ of the excitation regions (in the AGN rest frame) for both forward and reverse shocks
is indicated.
Under the assumption that the plasma is fully relativistic in all zones, 
the resulting shocked plasma speed is $v_p'=0.64c$, and the reverse and forward shock speeds are
$v_r'=0.37c$ and $v_f'=0.86c$, respectively.
\label{spacetime}}
\end{figure}

We consider a collision between a shock moving down the jet at speed $v'_s$
and a stationary Mach disk (see Fig.~\ref{jet}).
All the primed quantities are  evaluated in the AGN rest frame,
whereas all the plasma rest frame quantities are unprimed.
Prior to collision, the moving shock wave approaches stationary Mach disk, which results in 
gradual shrinking of zone (3) and eventually a head-on collision between
the moving shock and the Mach disk.
The reverse and forward shocks (separated by a contact discontinuity) that result from
the collision are shown in the space-time diagram in Fig.~\ref{spacetime}.
Both shocks propagate downstream in the frame of the AGN, but move in opposite
directions in the shocked plasma rest frame.
Shocked plasma occupies zones $(r)$ and $(f)$.
Following the collision that occurs at time $t'=0$, the forward shock expands into the dense medium accumulated
behind the Mach disk [zone $(4)$] until it reaches the rarefaction region [zone $(5)$] 
at time $t'_{cf}$.
Similarly, the reverse shock expands into zone $(2)$, which consists of the material swept up 
by the moving shock prior to collision.
We assume that the acceleration in the reverse zone effectively stops 
once the reverse shock reaches the contact discontinuity between $(1)$ and $(2)$ at time $t'_{cr}$.
It should be noted that the forward and reverse shocks interacting with the contact discontinuities
or the rarefaction behind the Mach disk may set up secondary shocks and/or waves that could complicate the
system further \citep[see][]{gom97}; we ignore these affects for the sake of simplicity.

The properties of the shocked plasma can be found by using the shock jump conditions
for a relativistic gas \citep[see][]{ani89}, which are based on the conservation of
matter, momentum, and energy across the shock front.
In this section we set the speed of light $c$ to unity for convenience.
For 1D flow, the conserved quantities reduce to
\begin{eqnarray}
\rho\Gamma(v-v_{\Sigma})=\mbox{const},\nonumber\\
\Gamma^2(v-v_{\Sigma})(e+p)+pv_{\Sigma}=\mbox{const},\\
\Gamma^2(v-v_{\Sigma})(e+p)v+p=\mbox{const}.\nonumber
\end{eqnarray}
These relations connect proper density $\rho$, pressure $p$, internal energy $e$, 
and plasma speed $v$ in units of $c$
(the corresponding Lorentz factor $\Gamma=1/\sqrt{1-v^2}$)
on both sides of the shock front, which moves at speed $v_{\Sigma}$.
They can be applied in any reference frame.
If conditions upstream of the shock as well as the speed of the shock $v_{\Sigma}$
are known, the shocked plasma parameters can be calculated provided that the equation
of state is specified.
We assume a fully relativistic plasma, in which case the equation of state is $p=e/3$.

Let us assume that plasma properties $(\rho_3,p_3,v_3)$  in zone $(3)$ are known.
In order to make the problem tractable, we approximate that velocity, pressure, and density 
are constant within each zone.
Since the speeds of both the moving shock and the Mach disk are known ($v'_M=0$ for the Mach disk),
we can therefore determine all the plasma parameters in zones $(2)$ and $(4)$ as described below.
However, the speeds of the forward and reverse shocks are unknown, so the conditions
in zones $(r)$ and $(f)$ must be determined simultaneously by appealing to the constancy of 
the plasma speed $v'_p$ and pressure $p$ across the contact discontinuity between $(r)$ and $(f)$.
Applying shock jump formulas to both zones in the plasma rest frame gives six equations for six unknowns:
$\rho_r$, $\rho_f$, $v_r$, $v_f$, $p$, and $v'_p$.
After some substitutions we find the following equation:
\begin{equation}\label{vp}
\frac{1-3v_rv_2}{p_2(1+v_rv_2)}=\frac{1-3v_fv_4}{p_4(1+v_fv_4)},
\end{equation}
where $v_f=(\sqrt{v_4+3}-v_4)/3$ and $v_r=(\sqrt{v_2+3}-v_2)/3$ are 
the forward and reverse shock speeds in the plasma rest frame.
In this reference frame the shock speeds cannot exceed $1/\sqrt{3}$
but they are also greater than $1/3$ for a fully relativistic plasma.
Eq.~(\ref{vp}) can be solved for $v'_p$ by applying relativistic
transformation formulas,
\begin{equation}
v_4=\frac{v'_p-v'_4}{1-v'_pv'_4}\quad\mbox{and}\quad
v_2=\frac{v'_2-v'_p}{1-v'_2v'_p},
\end{equation}
that connect the AGN rest frame speeds (primed quantities) and the plasma rest frame speeds (unprimed).
Since $v'_4<v'_p<v'_2$ in the AGN rest frame, speeds $v_2$, $v_4$, $v_f$, $v_r$ are defined to be positive
in the above formulas 
for convenience despite the fact that the shocks move in opposite directions in the plasma rest frame.

Although Eq.~(\ref{vp}) depends on $p_2$ and $p_4$, these quantities scale as $p_3$ for a fully relativistic plasma
and therefore the plasma speed $v'_p$
only depends on the quiescent flow speed $v'_3$ and the moving shock speed $v'_s$.
In general, Eq.~(\ref{vp}) must be solved numerically, but an important analytical solution
can be obtained if $p_2=p_4$.
This is equivalent to stipulating that in the rest frame of the quiescent flow [zone (3)]
the speeds of the moving shock and the Mach disk are equal.
This condition is satisfied if $v'_s=2{v}'_3/(1+{v'}^2_3)$ or 
$\Gamma'_s={\Gamma'}^2_3(1+{v'}^2_3)$.
The resulting shocked plasma speed is $v'_p=v'_3$, which also gives $\rho_r=\rho_f$ and
$v_r=v_f$.
Indeed, if the colliding shock speeds are equal in the rest frame of the quiescent flow 
then the shocked plasma will remain stationary in this frame while
the speeds of the reflected shocks will be equal.

\begin{figure}
\epsscale{0.5}
\plotone{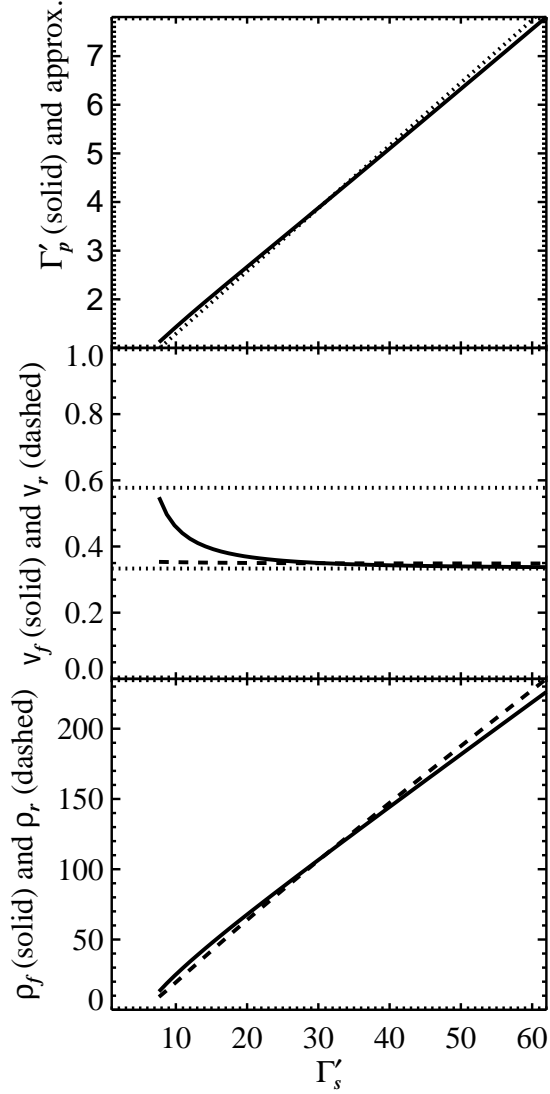}
\caption{Shocked plasma parameters as a function of the moving shock Lorentz
factor $\Gamma_{s}'$ for $\rho_3=1$, $p_3=1$, and $\Gamma_3'=4$;
the plasma is assumed to be fully relativistic.
Approximate solution for the shocked plasma Lorentz factor (dotted line) is shown in
top panel.
Speeds of $v=1/\sqrt{3}$ and $v=1/3$ are indicated by dotted lines in the middle panel for comparison
(speed of light $c=1$).
Gradients along the jet axis may change the ratio of densities in zones (r) and (f).
 \label{vpplot}}
\end{figure}

Fig.~\ref{vpplot} presents the shocked plasma Lorentz factor as a function
of the moving shock Lorentz factor (top panel) for a constant value $\Gamma_3'=4$
and $\rho_3=1$, $p_3=1$.
We have used the equation of state for a fully relativistic plasma in all zones.
The minimum possible shock speed of $v_s=1/\sqrt{3}$ in the quiescent plasma
rest frame, combined with the bulk Lorentz factor of the quiescent plasma,
 give
the lower limit on possible values of the moving shock Lorentz factor, $\Gamma_{s,min}'\approx7.6$.
The condition $p_2=p_4$ is satisfied when $\Gamma_s'\approx31$ for the chosen
value $\Gamma_p'=4$.

Using the analytical solution $v'_p=v'_3$ discussed in the previous paragraph
allows one to find an approximation,
\begin{equation}\label{vpapprox}
\Gamma_p'\sim\frac{\Gamma_s'}{\Gamma_3'(1+{v_3'}^2)},
\end{equation}
that gives an exact solution for the $p_2=p_4$ case and
does not deviate much from the exact solution when $\Gamma_s'\sim{\Gamma_3'}^2(1+{v_3'}^2)$
and $\Gamma_p'\gg1$.
Approximation~(\ref{vpapprox}) is shown in the top panel of Fig.~\ref{vpplot} (dotted line).
It should be noted that extremely fast relativistic shocks are required 
to achieve moderate shocked plasma speeds $v_p'$, e.g., $\Gamma_s'=31$ produces
only $\Gamma_p'=4$.

Reverse and forward shock speeds and plasma densities in the shocked
plasma rest frame are shown in the middle and bottom panels of Fig.~\ref{vpplot}.
An interesting result is that the value of the reverse and forward densities $\rho_r$ and $\rho_f$
are similar for most values of $\Gamma_s'$ and $\Gamma_3'$ (see bottom panel of Fig.~\ref{vpplot}). 
This is to be expected when $p_2\sim p_4$.
To understand why the densities are still close even when $p_2$ is much different from $p_4$,
consider that the final densities $\rho_r$ and $\rho_f$ result
from two consecutive amplifications from the starting value of $\rho_3$.
The two amplifications that result in the final density $\rho_f$ occur at the Mach disk and the forward shock,
and at the moving shock and the reverse shock for $\rho_r$.
For example, $\Gamma_s'=9$ is well below 
the critical value of  $31$ that gives $\rho_r=\rho_f$ (Fig.~\ref{spacetime}).
Thus, the density amplification caused by the moving shock is small compared with that caused by the Mach disk.
Since the moving shock is relatively weak, its collision with the Mach disk will
slow down the plasma to a mildly relativistic speed (see Fig.~\ref{spacetime}).
Therefore, the amplification by the forward shock will be small whereas
that of the reverse shock will be relatively large because it encounters
fast plasma from zone $(2)$.
For this reason, the forward and reverse densities resulting from these double amplifications are similar.
The same reasoning applies to the case when the moving shock is stronger than
the Mach disk, resulting in faster moving plasma.
The amplification of the transverse (to the flow) component of 
dynamically insignificant magnetic field should exhibit similar behavior.

We note that the above is strictly true only if quiescent-flow plasma in zone $(3)$ 
is homogeneous, without any gradients along the jet axis.
Since the jet may be expected to converge somewhat at the position of the Mach disk,
$\rho_3$ must be larger at the Mach disk.
If properly taken into account, this could result in higher values of the forward density $\rho_f$
and magnetic field strength $B_f$.

\section{Structure of the emission region}
\label{model.light}
It is clear that once the profiles of the individual flares are resolved,
a more detailed model of the emission region 
that takes into account the excitation structure, geometry of the region,
and the finite speed of light propagation must be used to reproduce observed flares.
It is usually assumed that the relativistic electrons are injected
uniformly in space within the source.
The mechanism that can provide this type of injection is often left unspecified,
and it is indeed difficult to conceive of a physical process that can change conditions 
simultaneously across the emission region.
We therefore adopt a scenario in which changes are transmitted by an excitation front,
which we assume to be a shock.
Because of highly relativistic motion of the source in the observer's frame,
one needs to take into account the light travel time and relativistic
effects that may dramatically influence time scales, frequencies, and flux levels
of the observed radiation.
Furthermore, in the SSC model another type of light-travel time effect must be considered.
The inverse Compton scattering at a particular location within the confines of the excited region
is likely to involve seed photons produced throughout the region.
Many of these seed photons travel across a significant fraction of the excited region.
These travel times may be
comparable to the characteristic time scale of structural changes of the source
excited by a relativistic shock wave moving at almost the speed of light.

\subsection{Excitation structure and geometry}
In this section we discuss the geometry of the source,
the excitation structure introduced by a passing shock wave, and
the effect of light travel delays on the internal structure
as viewed from different locations, for example, as seen by a distant observer.
In our model, we approximate that the excitation region has a cylindrical shape
with the axis of the cylinder coinciding with the axis of the jet.
Its transverse extent is determined by the size of the Mach disk,
which could be much smaller than the cross-section of the jet itself.
The region is filled uniformly with relativistic electrons gyrating in
a mostly turbulent magnetic field.

\begin{figure}
\epsscale{1.0}
\plotone{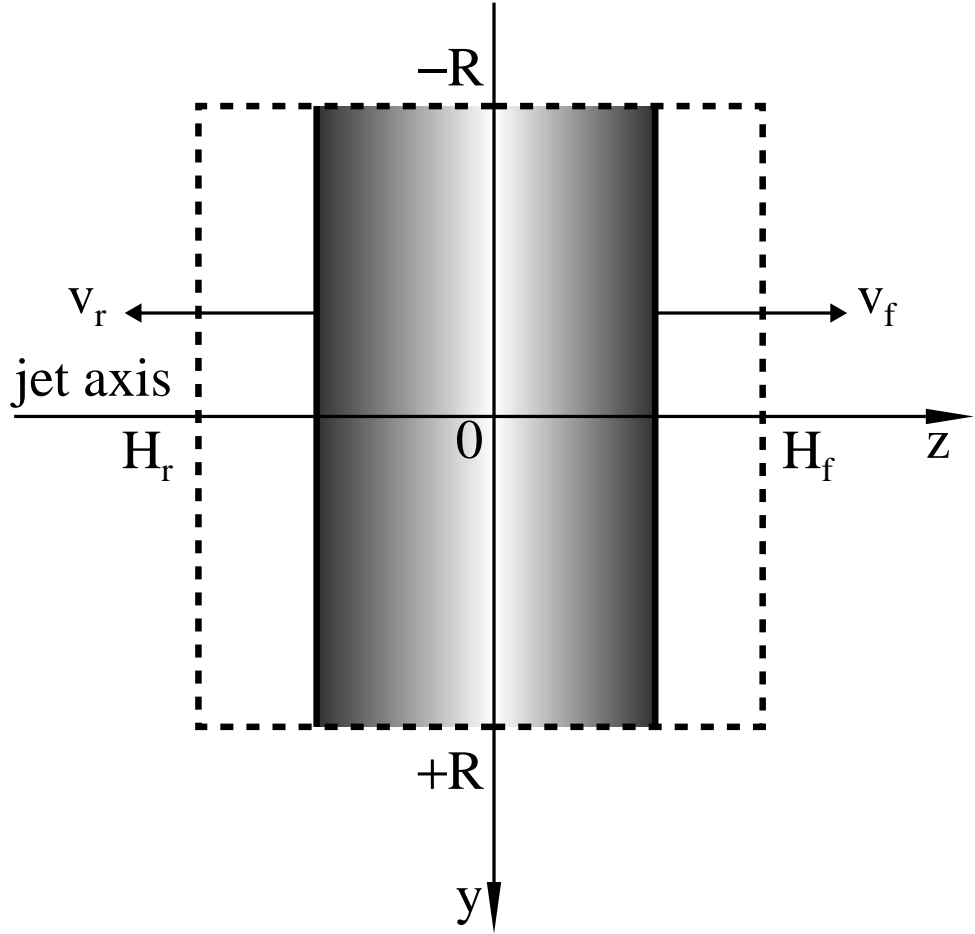}
\caption{
A snapshot of the electron excitation structure  while both forward
and reverse shocks (vertical solid lines) are inside the
excitation zones defined by $H_f$ and $H_r$.
The shock fronts move in opposite directions in the rest frame of
the emitting plasma at speeds $v_f=0.45\mbox{c}$ and $v_r=0.495\mbox{c}$.
The surface $z=0$ corresponds to the contact discontinuity
between the forward and reverse zones.
The emitting electrons are located between the forward and reverse shock fronts;
gray scale corresponds to the maximum energy of electron,
which is highest at the shock fronts.
\label{shock}}
\end{figure}

A snapshot of the excitation structure after the collision is shown in Fig.~\ref{shock}.
The electrons are accelerated in a very narrow region at the shock wave fronts,
which are assumed to be much smaller than the dimensions of the source.
As soon as electrons leave the acceleration region, their energy steadily decreases
due to radiative energy losses.
Although the density of electrons does not change, the maximum
energy of electrons declines, establishing a gradient in the electron
distribution across the source.
The farther away from the shock front a given location is,
the lower is the maximum energy in the distribution.
The excitation structure follows the progression of shocks through the 
forward and reverse regions, which are assumed to be moving at constant speeds.

Because of intrinsic similarity between the forward and reverse regions,
here and in the following sections we mostly consider the formulas for the forward emission region.
We drop the subscript $f$ unless both regions must be considered.
Analogous expressions for the reverse region can be 
obtained straightforwardly by inverting the $z$-axis.
The evolution of the distribution of electrons is found by solving the equation
\begin{equation}\label{Nevolve}
\frac{\partial{N}}{\partial{t}}+\frac{\partial{\left(\dot{\gamma}N\right)}}{\partial{\gamma}}=0
\end{equation}
with the initial distribution $N(\gamma)=N_0\gamma^{-s}$ and $\gamma_{min}<\gamma<\gamma_{max}$.
We assume that there are no sources or sinks of electrons in the region
between the acceleration zone and outer edge of the volume of excited electrons.
We also neglect adiabatic losses since the expansion of the emitting volume is
insignificant during relatively fast flares caused by collisions with the Mach disk.
The solution depends on time $t$ and position along the jet axis $z$.
The energy loss function $\dot{\gamma}$ depends on $\gamma$
and the average magnetic field
strength $B$ if synchrotron losses provide the dominant contribution: $\dot{\gamma}=-\gamma^2/t_1$,
where $t_1=7.73\times10^8\,\mbox{s}\,/(B/1\,\mbox{G})^2$.
If inverse Compton losses are significant, then $\dot{\gamma}$ will also depend on the
properties of the seed photon field.
In the case of the SSC model, the inclusion of Compton losses significantly
complicates the problem, since the seed emission itself depends on the distribution
of electrons.
However, Eq.~(\ref{Nevolve}) can be solved independently of the radiation field structure 
if SSC losses are neglected.
For constant magnetic field, the solution is then
\begin{equation}\label{gammamax}
\gamma_{max}(\Delta{}t)=\frac{\gamma_{max}}{1+\gamma_{max}\Delta{t}/t_1}
\end{equation}
and
\begin{equation}
N(\gamma)=N_0\gamma^{-s}\left(1-\gamma\Delta{t}/t_1\right)^{s-2},
\end{equation}
where $\Delta{}t=t-z/v$ is the time at position $z$ since the shock wave front passed it
and $v$ is the speed of the forward shock.
The darkest shade of gray in Fig.~\ref{shock} corresponds to $\Delta{}t=0$ since it is
located at the current position of the shock wave front, where the electron energy distribution
extends up to $\gamma_{max}$. 
For times and positions within the source corresponding to $\Delta{}t>0$,
the shape of the distribution is different from the original if $s\neq2$
and the maximum Lorentz factor of the distribution is less than $\gamma_{max}$.
The value of $\gamma_{max}(\Delta{}t)$ begins to deviate significantly 
from the initial value when $\Delta{}t\sim t_1/\gamma_{max}$,
which is about $10^6\,\mbox{s}$ for $B=1\,\mbox{G}$ and $\gamma_{max}=10^3$.
For example, for a source of size $\sim0.01\,\mbox{pc}$
(light crossing time  $\sim10^6\,\mbox{s}$) only the subset of
electrons with Lorentz factors $\gamma\ll10^3$ may be considered homogeneous
across the source.
For larger source dimensions the effect of inhomogeneity is even more severe.

Light travel effects modify the appearance of the excitation structure and, therefore,
the observed flux.
It is natural to consider the internal structure in the rest frame of the shocked plasma, 
since it is the most natural frame for calculating the emission.
All quantities discussed in this section are defined in the plasma
rest frame.

\begin{figure}
\plotone{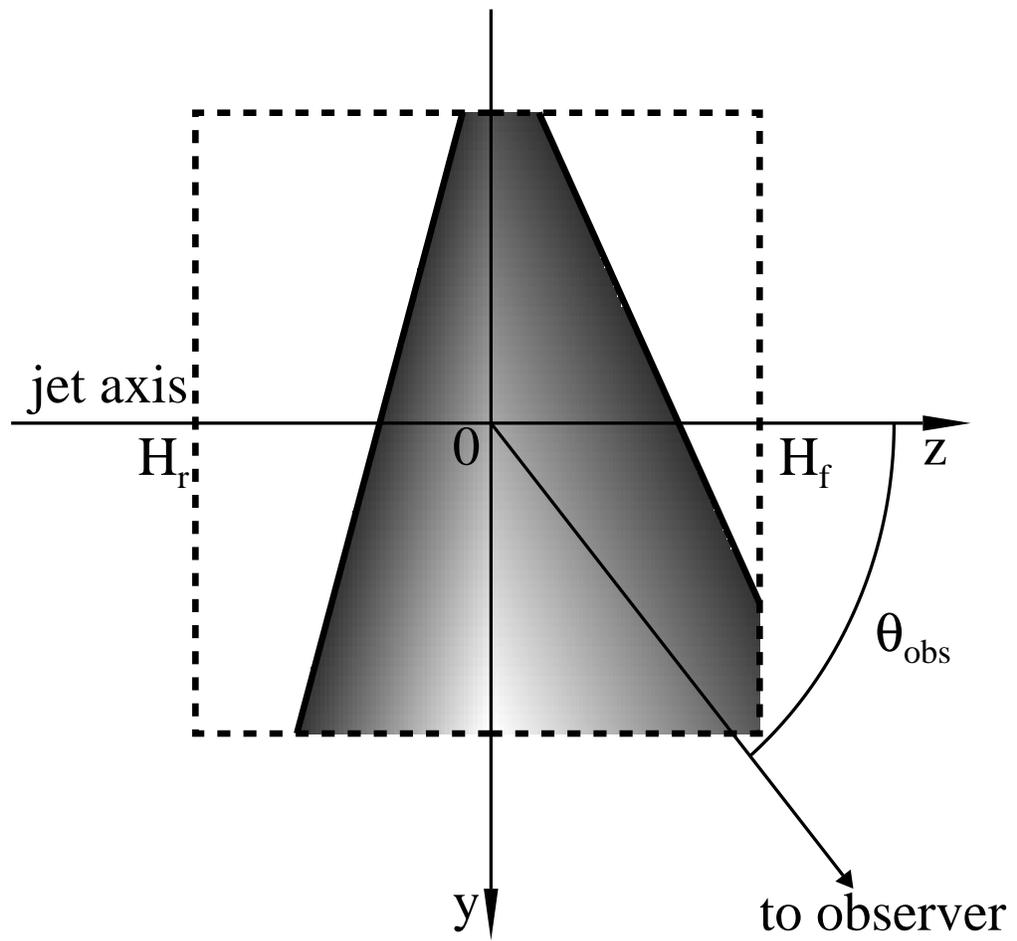}
\caption{A snapshot of the apparent excitation structure of the source as seen by a distant
observer in the plasma rest frame; the line of sight subtends  an angle $\theta_{obs}$ to the z-axis.\label{apshock}}
\end{figure}

The orientation of the apparent shock  fronts and the internal structure
of the source
for an observer in the plasma rest frame who is located in the $y$-$z$ plane 
at an angle $\theta$ to the $z$-axis are presented in Fig.~\ref{apshock}.
The equation that defines the position of the apparent shock front is
\begin{equation}\label{fapshock}
(1- v\cos{\theta_{obs}}/c)z=(y-R)v\sin{\theta_{obs}}/c+v t_{obs},
\end{equation}
where $t_{obs}$ is the time of observation.
We define $t_{obs}=0$ as the time when the first radiation from the source is detected by the observer.
There is no direct relation between $t_{obs}$ and the local emission time $t$  because
at any given time $t_{obs}$ the observer detects radiation 
from various locations in the source produced at different times. 
As illustrated in Fig.~\ref{apshock}, the radiation coming from the far side of source
(top of the sketch) is delayed with respect to that produced at the near side of the source
owing to light travel time.

The apparent shock position determines which portion of the source will be
visible to the observer at the time of observation $t_{obs}$.
Every  point in the source behind the apparent shock front contributes
to the observed radiation at time $t_{obs}$ but at different local times $t$.
A range of physical shock positions in the source corresponds to a single time $t_{obs}$.
For any given position $(x,y,z)$ and time $t_{obs}$ the local emission time $t$
is given by
\begin{equation}\label{ttobs}
t=t_{obs}+[(y-R)\sin{\theta_{obs}}+z\cos{\theta_{obs}}]/c.
\end{equation}
The last two terms in this equation determine the shift along the line of sight of the local point $(x,y,z)$
with respect to $(0,R,0)$, which by definition has no delay.
Once the local time of emission is found, we can determine $\Delta{}t$ and  the
local distribution function of electrons, which is then used to calculate the emissivity.

The apparent crossing time $t_{ac}$ is the time when the shock wave front appears to have left the source
and is found by substituting $z=H$ and $y=-R$ into Eq.~(\ref{fapshock}).
This gives
\begin{equation}\label{apcrossing}
t_{ac}=2R\sin{\theta_{obs}}/c+H(1/v-\cos{\theta_{obs}}/c).
\end{equation}
Since the speed of the shock front in the shocked plasma frame cannot exceed $c/\sqrt{3}$,
the apparent crossing time in the plasma rest frame can never  be significantly
smaller than the physical crossing time $H/v$
despite relativistic speeds of colliding shocks.
Higher energy electrons with fast decay times occupy a relatively thin region behind the
apparent shock front and, therefore, the emission that they produce 
traces the geometry of the excitation region along the path of the apparent shock front.
The corresponding flare peaks at half the apparent crossing time $t_{ac}/2$.
On the other hand, lower energy electrons completely fill up  the portion of the source that
has been excited by the shock.
Therefore, these electrons will produce emission that traces the entire volume of source
and the associated flare will peak at the crossing time $t_{ac}$.

Synchrotron emission from the forward and reverse regions of the source 
can be calculated  independently of each other at  frequencies
above the synchrotron self-absorption frequency.
If the angle of observation $\theta_{obs}$ is different from $90^{\circ}$, then one of the 
regions, reverse or forward, will block parts of the adjacent region in the optically thick case.

\subsection{Light Travel Time Effects for SSC Emission} 
In the SSC model, a population of relativistic electrons that emit synchrotron radiation
also up-scatters some of these synchrotron photons, thereby giving rise to  inverse
Compton emission.
Electrons at every local position within the source scatter the photons that are produced locally 
as well as those arriving from other portions of the source.

Because of this expenditure of energy, the average and maximum energy of the
population of electrons decays with time.
The intensity of synchrotron radiation produced locally therefore also decreases.
However, even when the local electron population has decayed substantially the synchrotron
emission arriving from other parts of the source may still contain plenty of
high energy photons to drive inverse Compton emission.

\begin{figure}
\plotone{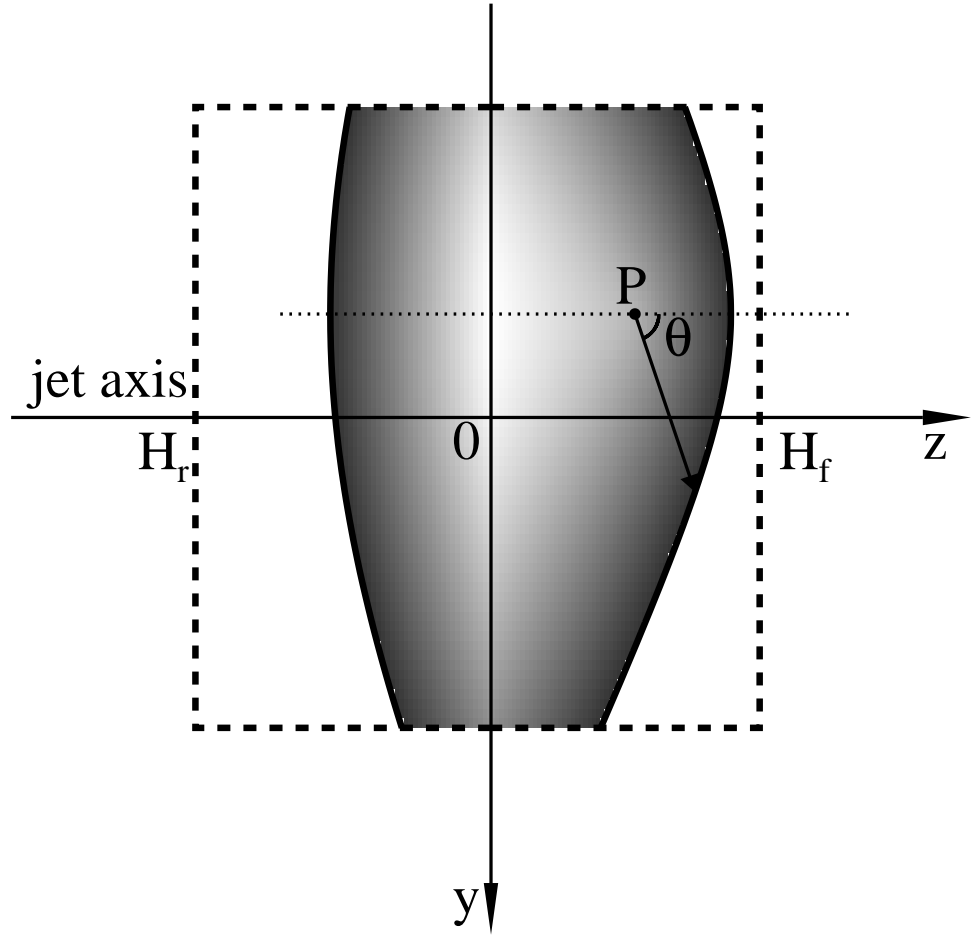}
\caption{\label{inapshock}
Internal apparent shock orientation (solid lines) seen at an arbitrarily chosen point $P$ in the forward zone at some
time after the shock front passes it. 
The distance between point $P$ and the apparent shock position at an arbitrary angle $\theta$
to the $z$-axis depends on the light travel time from this position to point $P$
and the speed of the shock front.}
\end{figure}

At an arbitrary point $P=(x_P,y_P,z_P)$ in the forward region
the appearance of the shock wave front, which we call the internal apparent shock front,
is non-planar. 
For time $t_P>z_P/v_f$ 
the distance from $P$ to the internal apparent shock front in direction
$\theta$ (see Fig.~\ref{inapshock})
is given by
\begin{equation}\label{apshockf}
r=\frac{v_f{}t_P-z_P}{v_f/c+\cos{\theta}}
\end{equation}
for the forward shock front and
\begin{equation}\label{apshockr}
r=\frac{v_r{}t_P+z_P}{v_r/c-\cos{\theta}},
\end{equation}
for the reverse shock front in the adjacent region.
Both reverse and forward regions will contribute synchrotron photons for the inverse Compton scattering.
Thus, in general it is necessary to consider the incident emission originating from both region.

To calculate the inverse Compton emission coefficient at point $P$ one
needs to find the synchrotron radiation field at this location, i.e., the intensity of
the synchrotron emission as a function of direction and frequency.
The synchrotron emission comprises light produced at different locations at different times
within the excitation region defined by the apparent positions of the shock fronts.
The time of emission from a specific location $(r,\phi,\theta)$ is $t=t_P-r/c$, whereas
the shock passage time at this location is $|z|/v_{r,f}$,
where $z$ is the Cartesian coordinate of the location.
Therefore, one determines the time interval $\Delta{}t=t-|z|/v_{r,f}$ between the emission
and shock front passage for this location.
The interval $\Delta{}t$ determines the local distribution function,
and hence the emitted synchrotron radiation,  at this location.

\section{Calculation of Emission}
\label{model.emission}
As described above, for calculating the flux in the plasma rest frame we approximate 
that the emission region has a cylindrical geometry and is filled with relativistic
electrons excited by a shock wave passing through it at a constant speed.
The density of relativistic electrons and 
magnetic field strength are taken to be constant, although the distribution of
electrons changes according to the energy losses that they experience.
The magnetic field is randomly oriented except for a possible small uniform component.
A power-law energy spectrum of electrons is injected at the shock wave front.
We use approximate expressions for the synchrotron emission coefficient as described in the
following sections.
Once the flux in the plasma rest frame is found, the observed quantities are
calculated by using the transformation formulas discussed below.
The following set of input parameters for each emission region is used:
geometrical dimensions of the cylinder $R$ 
(the same for both reverse and forward regions) and $H$; 
plasma number density $n$;
magnetic field strength $B$; distribution of electrons specified by parameters 
$\gamma_{min}$, $\gamma_{max}$, and $s$; shock speed in the plasma rest frame  $v$.
These are complemented by the  
angle between the line of sight and jet axis $\theta_{obs}$; 
bulk plasma speed $\Gamma_p'$; and redshift $z$, which defines the distance to the blazar.
The output of the model is the flux density as a function of frequency and time.

\subsection{Synchrotron Emission}
We use radiative transfer equation and the method of characteristics to calculate 
the intensity and observed flux given the synchrotron emission and absorption coefficients.
Following \citet{pac70}, the synchrotron emission coefficient is
\begin{equation}
j_{\nu}(\Delta{t})=c_3B\int{}N(\gamma,\Delta{t})F(x)\,d\gamma,
\end{equation}
where $x=(\gamma_{\nu}/\gamma)^2$,
$\gamma_{\nu}=\sqrt{\nu/(c_1B)}$ is the Lorentz factor of an electron 
(with pitch angle equal to the mean)
that emits at critical frequency $\nu$, 
$c_1=2.8\times10^6$, and 
$c_3=1.87\times10^{-23}$ in cgs units.
The integral is taken from ${\gamma_{min}(\Delta{t})}$ to ${\gamma_{max}(\Delta{t})}$ where
$\Delta{t}$ is the time since the shock front passage at the current location.
The  function $F(x)=x\int^{\infty}_{x}K_{5/3}(z)\,dz$,
where $K_{5/3}$ is a Bessel function of the second kind, can be replaced with its approximation
\citep{pac70}.
After rejecting insignificant terms, we obtain
\begin{equation}
j_{\nu}(\Delta{t})\approx\frac{c_3BN_0}{\gamma_{\nu}^{s-1}}
\exp{\left(-\left[\gamma_{\nu}/\gamma_{max}(\Delta{t})\right]^2\right)}.
\end{equation}
The approximation is different from the exact formula by a factor of order unity that
depends weakly on $s$, $\nu$, and $\Delta{t}$.
Assuming that the synchrotron emission by electrons provides the dominant cooling, we
can substitute solution (\ref{gammamax}) into the above equation.
To facilitate the ease of analytical considerations the resulting expression can be further simplified,
which gives the final expression used in calculations and analytical considerations:
\begin{equation}
j_{\nu}(\Delta{t})=\frac{c_3BN_0}{\gamma_{\nu}^{s-1}}
\exp{ (-\nu/\nu_{max})     }
\exp{ (-\Delta{}t/t_{\nu}) },
\end{equation}
where $\nu_{max}=c_1B\gamma_{max}^2$ and $t_{\nu}=t_1/\gamma_{\nu}$ is the characteristic
decay time of electrons owing to synchrotron losses.
A similar analysis can be carried out for the absorption coefficient, with a final result
\begin{equation}
\alpha_{\nu}(\Delta{t})=\frac{c_3BN_0}{2\gamma_{\nu}^sm_e\nu^2}
\exp{ (-\nu/\nu_{max})     }
\exp{ (-\Delta{}t/t_{\nu}) },
\end{equation}
where $m_e$ is the mass of electron.
The ratio of the coefficients gives the source function, which is independent of time and position:
\begin{equation}
S_{\nu}=j_{\nu}/\alpha_{\nu}=2\gamma_{\nu}m_e\nu^2.
\end{equation}
This allows us to derive an analytical solution to the radiative transfer of the synchrotron emission.

\subsection{Radiative Transfer}
\label{radtrans}
The flux density measured by an observer at distance $D_0$ in the plasma rest frame is
\begin{equation}\label{anfirst}
F_{\nu}(t_{obs})=\frac{1}{D_0^2}\int\int{}I_{\nu}(t_{obs},\xi,\eta)\,d\xi\,d\eta.
\end{equation}
It is evaluated by integrating intensity observed at time $t_{obs}$ 
over the projection of the source on the plane of the sky.
The intensity along a given path defined by $(\xi,\eta)$ 
is found by solving the time-dependent radiative transfer equation,
\begin{equation}\label{trans}
\frac{1}{c}\frac{\partial{I_{\nu}}}{\partial{t}}+\frac{\partial{I_{\nu}}}{\partial{\zeta}}=
-\alpha_{\nu}I_{\nu}(t,\zeta)+j_{\nu}(t,\zeta),
\end{equation}
where $\zeta$ is the coordinate along the line of sight.
The transfer equation is integrated along a path through the excited region
determined by the position of the apparent shock front  at time $t_{obs}$ and the geometry of the source.
Using the characteristics $t(\zeta)=t_{obs}+\zeta/c$, which describe a set of space-time positions  
through the source that produce the emission arriving at the observer at time $t_{obs}$,
we can convert partial differential equation (\ref{trans}) into an ordinary differential equation
that can be integrated with the above approximations for the emission/absorption coefficient.
The result is
\begin{equation}
I_{\nu}(t_{obs},\xi,\eta)=S_{\nu}(1-\exp{(-\tau_{\nu})}),
\end{equation}
with the optical depth defined by
\begin{equation}
\tau_{\nu}(t_{obs},\chi,\eta)=\alpha_{\nu}(0)\Delta{\zeta}T_{\nu}(\chi)\exp{(-\Delta{t}/t_{\nu})}.
\end{equation}
Here, $T_{\nu}(\chi)=(\exp{\chi}-1)/\chi$ and
\begin{equation}\label{anlast}
\chi=(\cos{\theta_{obs}-v/c})\Delta{\zeta}/(v{}t_{\nu}).
\end{equation}
The length $\Delta{\zeta}=\zeta_{max}-\zeta_{min}$ is the geometric thickness of the source
along the given path through it.
By combining Eqs.~\ref{anfirst}-\ref{anlast} with the  expression for $\Delta{t}$,
\begin{equation}
\Delta{t}
=
t_{obs}
+
\eta\sin{\theta_{obs}}
/
v
+
\left(
  1/c
  -
  \cos{\theta_{obs}}
  /
  v
\right)
\zeta_{min},
\end{equation}
one can obtain an analytical expression for the evolution of the observed synchrotron spectrum
if the geometry of the source is sufficiently simple.
For a cylindrical source, the expression for $\Delta{\zeta}$
is greatly simplified if the direction toward the
observer coincides with the axis of the jet.
In this case the solution is
\begin{equation}\label{asol0deg}
F_{\nu}(t_{obs})=\frac{\pi{}R^2}{D_0^2}[1-\exp(-\tau_{\nu})]S_{\nu},
\end{equation}
with
\begin{equation}\label{asol0deg.tau}
\tau_{\nu}=\alpha_{\nu}(0)H\exp(-t_{obs}/t_{\nu})\frac{\exp(\min{\{t_{obs},t_{ac}\}}/t_{\nu})-1}{t_{ac}/t_{\nu}},
\end{equation}
where $t_{ac}=(c/v-1)H/c$ for the forward region.

\subsection{Self-Compton Emission}
Similar to the synchrotron case, the inverse Compton flux density is obtained by integrating
intensity over the extent of the source on the plane of the sky.
Since we assume that the inverse Compton emission is optically thin at all frequencies
of interest, the intensity is the integral of the inverse Compton emission coefficient
along a given path toward the observer, which gives
\begin{equation}
F_{\nu}^C(t_{obs})=\frac{1}{D_0^2}\int\int\int{j_{\nu}^C(t,\xi,\eta,\zeta)}\,d\xi\,d\eta\,d\zeta,
\end{equation}
where $t(\zeta)=t_{obs}+\zeta/c$ takes into account the light travel effects.

The inverse Compton emission coefficient
\begin{equation}\label{compcoef}
j_{\nu}^C
=
\int d\Omega_{i}
\int d\gamma{}\,
  N(\gamma)
\int d{\nu_i}\,
\frac
{
h\nu
}
{
h\nu_i
}
I_{\nu_i}(\vec\Omega_i)
\psi(\nu_i,\nu,\vec\Omega_i,\vec\Omega_f,\gamma)
\end{equation}
depends on the local distribution of electrons $N(\gamma)$, which is  assumed to be isotropic,
and the incident synchrotron radiation $I_{\nu_i}(\vec\Omega_i)$ that provides the seed photons at different
frequencies $\nu_i$.
The redistribution function $\psi$ determines the fraction of incident radiation at initial frequency $\nu_i$
in initial direction $\vec\Omega_i$ that is scattered by electrons with Lorentz factor $\gamma$
into final direction $\vec\Omega_f$ with final frequency $\nu$.
The final direction $\vec\Omega_f$ is constant and determined by the angle $\theta_{obs}$.
The limits of the integrals, as well as the expression for the redistribution function, 
are discussed in Appendix~\ref{appredist}.
The redistribution function that we use applies at all electron energies and photon frequencies allowed
by energy and momentum conservation.
It does not make any simplifying assumptions about the angular distribution of the incident radiation,
and is accurate in both Thomson and Klein-Nishina regimes.
If $\gamma{}h\nu_i\ll{}m_ec^2$ (Thomson limit)
and  $\nu_i\ll{}\nu_f$,
expression (\ref{compcoef}) for the Compton emission coefficient simplifies to that
derived by \citet{rey82}.

Similar to the calculation of the observed emission, the internal 
synchrotron seed photon intensity is calculated by solving the time-dependent radiative
transfer equation for each direction $\vec\Omega_i$.
The characteristic equation is $t(r)=t-r/c$, where $r$ is the distance
from the point at which the intensity is evaluated in the direction opposite
to $\vec\Omega_i$, i.e., the direction from which the radiation comes.
If the same approximations for the absorption and emission coefficients are
employed, the resulting intensity is
\begin{equation}
\label{insynch}
I_{\nu_i}(t,x,y,z,\vec\Omega_i)=[1-\exp(-\tau_{\nu_i})]S_{\nu_i},
\end{equation}
where
\begin{equation}
\label{insynch.tau}
\tau_{\nu_i}=\alpha_{\nu_i}(0)L(\vec\Omega_i)T_{\nu_i}(\chi)\exp(-\Delta{}t/t_{\nu_i})
\end{equation}
and, for the incident emission from the forward region, 
$\chi=(v_f/c-\Omega_{iz})L_f(\vec\Omega_i)/(v_f{}t_{\nu_i})$ and
$\Delta{}t=t-z/v_f$.
Here, $L_f(\vec\Omega_i)$ is the distance to the edge of the excited material
in the forward region and
$\Omega_{iz}$ is the projection of vector $\vec\Omega_i$ onto the $z$-axis.
For the incident emission originating in the reverse region, 
$\chi=(v_r/c+\Omega_{iz})L_r(\vec\Omega_i)/(v_r{}t_{\nu_i})$
and $\Delta{}t=t-z/(c\Omega_{iz})$.

In calculating the inverse Compton flux density we use the analytical solution for the incident synchrotron intensity
and perform all calculations on an adaptive grid that takes into account the electron excitation structure
and the internal synchrotron emission field, which  can be highly inhomogeneous and anisotropic.

\subsection{Observed quantities}
The model calculations are performed in the plasma rest frame where
a fiducial observer is located at  distance $D_0$ along a line that subtends an angle $\theta_{obs}$ to the $z$-axis.
The flux density thus calculated  must be converted to that
received in the observer rest frame (denoted by star)  by taking into
account all the relativistic, light travel, and cosmological effects.

The relation between the angle of observation in the plasma rest frame $\theta_{obs}$
and in the observer frame $\theta_{obs}^\star$ is 
\begin{equation}
\cos{\theta_{obs}^\star}=\frac{v_p'/c+\cos{\theta_{obs}}}{1+v_p'\cos{\theta_{obs}}/c}.
\end{equation}
For $\theta_{obs}=90^{\circ}$, the viewing angle in the observer rest frame 
can be found from $\cos{\theta_{obs}^\star}=v_p'/c$,  which gives $\theta_{obs}^\star\approx1/\Gamma_p'$;
this is the angle that maximizes apparent superluminal motion for a fixed speed $v_p'$.

The transformation formulas for frequencies, time, and flux are simplified with the use
of the Doppler factor, defined as 
\begin{equation}
\delta=\frac{1}{\Gamma_p'(1-v_p'\cos{\theta_{obs}^\star}/c)}.
\end{equation}
Thus defined, the Doppler factor, combined with cosmological time dilation,
gives the following relations:
\begin{equation}
\nu^\star=\nu\delta/(1+z), \quad t_{obs}^\star=(1+z)t_{obs}/\delta,
\end{equation}
and
\begin{equation}
F_{\nu}^\star=(D_0/D^\star)^2\delta^3F_{\nu},
\end{equation} where $D^\star$ is the luminosity distance to the blazar measured by the observer.
Accordingly, the apparent crossing time $t_{ac}^\star=(1+z)t_{ac}/\delta$ in the frame of the observer.

\section{Model Light Curves and Spectra}
\label{model.properties}
In this section we present the result of our simulations.
We first consider the light curves and spectra from a single region of emission.
Contributions from the adjacent region can be neglected if it is geometrically
thinner and/or contains lower density plasma and weak magnetic field.
We compare synchrotron and inverse Compton flares with respect to 
possible time delays between high and low energy emission.
We finish this section by presenting results of a double region simulation
and our modeling of variability from the quasar 3C~273.
All the quantities discussed here, except for the flare in 3C~273
will be given in the rest frame of the emitting plasma. 
Corresponding quantities in the observer's frame can be easily
calculated if the speed of the plasma, orientation of the jet, and source redshift are known.

\subsection{Synchrotron: Single emission region}
\begin{figure}
\epsscale{0.6}
\plotone{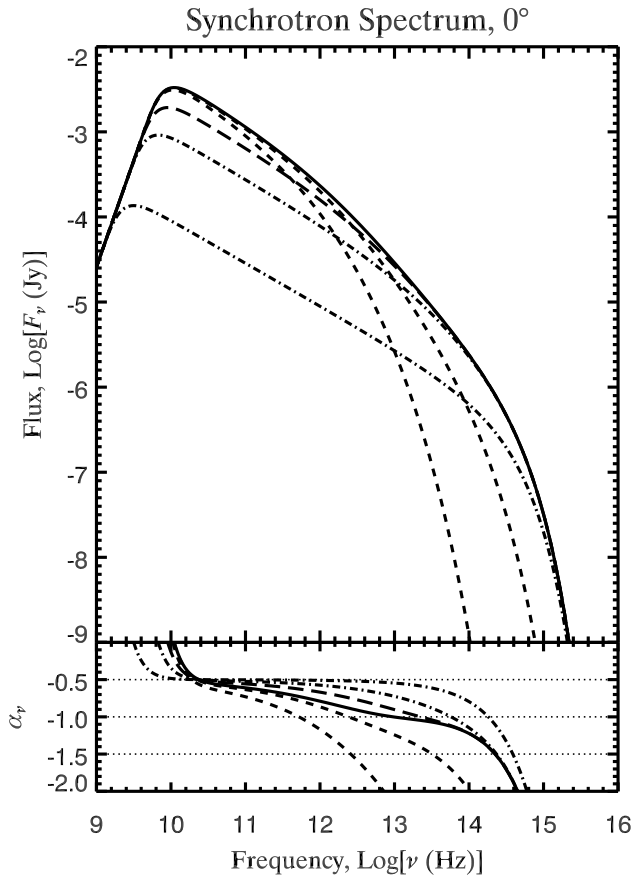}
\caption{The evolution of the synchrotron spectrum for $\theta_{obs}=0^{\circ}$
in the rest frame of the shocked plasma in the forward shock region
at times $t/t_{ac}$:
0.02, 0.2 (dot-dash), 0.5 (long dash), 1.0 (solid), 1.1, 1.5 (short dash).
The bottom panel indicates the evolution of the synchrotron spectral index $\alpha_{\nu}$, 
defined by $F_{\nu}(t_{obs})\propto\nu^{\alpha_{\nu}}$.
The following input parameters were used: $R=H=4\times10^6\,\mbox{s}$,
$v=0.34c$, $B=0.4\,\mbox{G}$, $n=10\,\mbox{cm}^{-3}$,
$\gamma_{min}=100$, $\gamma_{max}=2\times10^4$, $s=2$, $D_0=10^3\,\mbox{Mpc}$.
The synchrotron self-absorption frequency is $\nu_{t}=8\times10^9\,\mbox{Hz}$
and the maximum frequency is $\nu_{max}=5\times10^{14}\,\mbox{Hz}$.
\label{mssp.0}}
\end{figure}

\begin{figure}
\plotone{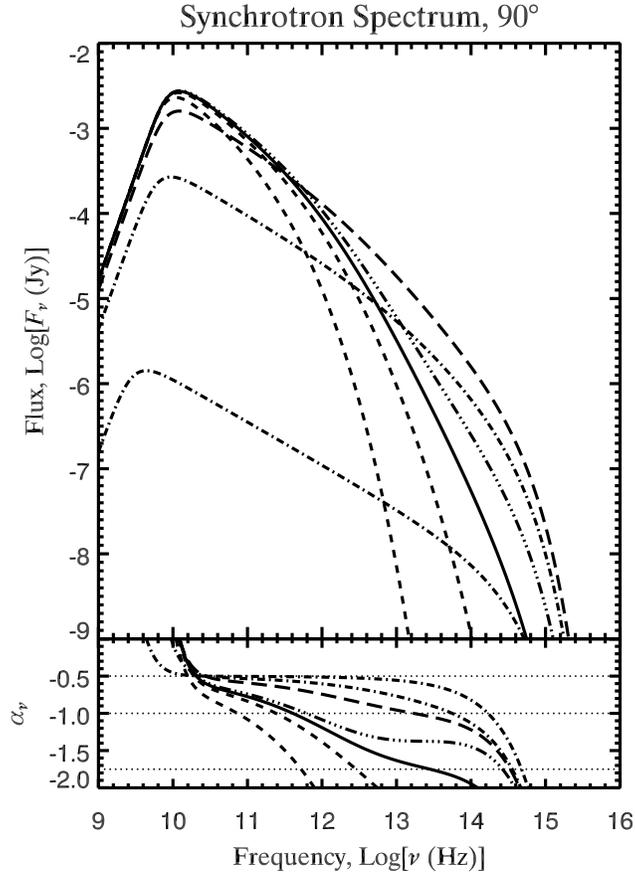}
\caption{The same as Fig.~\ref{mssp.0} but for $\theta_{obs}=90^{\circ}$,
which corresponds to observation angle $\theta^{\star}_{obs}\approx1/\Gamma'_p$.
The synchrotron self-absorption frequency is $\nu_{t}=10^{10}\,\mbox{Hz}$
and the maximum frequency is $\nu_{max}=5\times10^{14}\,\mbox{Hz}$.
More complex behavior of the spectrum is apparent because the shock wave 
traces a more complicated geometrical pattern for this viewing angle.
An additional spectrum at time  $t=0.94\,t_{ac}$ is shown with a triple-dot-dashed curve.
\label{mssp.90}}
\end{figure}

The evolution of the synchrotron spectrum during a flare is shown in Figs.~\ref{mssp.0} and~\ref{mssp.90}
for two representative viewing angles.
The shape of the synchrotron spectrum and its evolution can be roughly described 
by examining
three critical frequencies:
\begin{enumerate}
\item
The synchrotron self-absorption frequency $\nu_t$, at which the source becomes optically thick,
is found by solving the equation $\alpha_{\nu}(0)L=1$, where $L$ is the average geometrical thickness of
the source along the line of sight. 
The solution is 
\begin{equation}
\nu_t=\left(\frac{c_3N_0L}{2m_ec_1}\right)^{\frac{2}{s+4}}(c_1B)^{\frac{s+2}{s+4}}.
\end{equation}
During the flare the thickness $L$ is expected to grow as a result of the shock wave propagation through
the excitation region.
Even when the viewing angle $\theta_{obs}=90^{\circ}$, light travel effects
cause the apparent shock wave to propagate 
along the line of sight into, as well as across, the excitation region.
The growing $L$ will result in the increase of the self-absorption frequency $\nu_t$ until
the whole excitation region has been traversed by the shock.
\item
The expression for the break frequency $\nu_b$
at which the spectrum steepens owing to radiative energy losses 
depends strongly on the viewing angle.
Analytical solution~(\ref{asol0deg.tau}) of the transfer equation for $\theta_{obs}=0^{\circ}$
allows one to find the break frequency by solving the equation $t_{\nu}=\min\{t_{obs},t_{ac}\}$, which 
for $t_{obs}<t_{ac}$ gives 
\begin{equation}
\nu_b=c_1t_1^2/(t_{obs}^2B).
\end{equation}
Substituting $t_{obs}=t_{ac}$ and the input parameters used in simulations (see Fig.~\ref{mssp.0}) 
gives $\nu_b=6.3\times10^{11}\,\mbox{Hz}$.
At the beginning of the flare the spectral index, defined by $F_{\nu}(t_{obs})\propto \nu^\alpha$, 
is $-0.5$ at all frequencies
between the self-absorption and the drop-off frequencies but
it decreases quickly to $-1$ as the break frequency sweeps across the spectrum until
it reaches its minimum value of $6.3\times10^{11}\,\mbox{Hz}$ at $t_{obs}=t_{ac}$.
The spectral index steepens by $1/2$ above the break frequency due to the fact that
the depth of the actual emitting volume at $\nu>\nu_b$ is determined by the decay time
$t_{\nu}\propto\nu^{-1/2}$.
Similar reasoning may be used for different viewing angles, in which case Eq.~(\ref{apcrossing})
should be used to find the appropriate crossing time $t_{ac}$.
We stress that the break  in our simulated spectrum is 
only due to the frequency stratification of the emitting electrons.
\item
The exponential drop-off frequency $\nu_d$ remains 
equal to the critical frequency of the maximum-energy 
electrons, $\nu_{max}=c_1B\gamma_{max}^2$,
as long as the apparent shock
wave is in the excitation region. 
However, after the shock exits the excitation region the decrease in the drop-off frequency
mimics the decrease in the maximum electron energy (cf.~Eq.~(\ref{gammamax}) with $\Delta{}t=t-t_{ac}$).
\end{enumerate}

The assumption that the excitation region is a cylinder oriented along the axis of
the jet affects the behavior of the observed spectrum as a function of the viewing angle.
At viewing angle $\theta_{obs}=0^{\circ}$ the apparent surface of the shock propagates along the axis of
the cylinder and, therefore, the circular geometry of the source has no effect on the observed spectrum.
The evolution of the spectrum simply reflects the increase in depth through the source as the shock wave
propagates through it.
On the other hand, for viewing angle $\theta_{obs}=90^{\circ}$ the apparent surface of the shock front is tilted
away from the plane normal to the cylinder axis.
Thus, the apparent surface traces a roughly  circular shape for such a viewing angle;
this causes significant differences in the evolution of the observed spectrum
relative to the $\theta_{obs}=0^{\circ}$ case.
In particular, emission at the highest frequencies peaks roughly at half the apparent crossing time, since
the intersection between the excitation region and the shock front is largest at about that time.
The flux density will decrease significantly at these frequencies by the time the shock front exits the source.
At the crossing time $t_{obs}=t_{ac}$, the spectral index assumes a value of about  $-1.75$ at the highest frequencies
owing  to a unique combination of the cylindrical geometry and the orientation of the excitation structure in the source.
This stage is very brief and requires exact cylindrical geometry; hence, it is unlikely to be observed
but can be used to validate numerical calculations.

\begin{figure}
\plotone{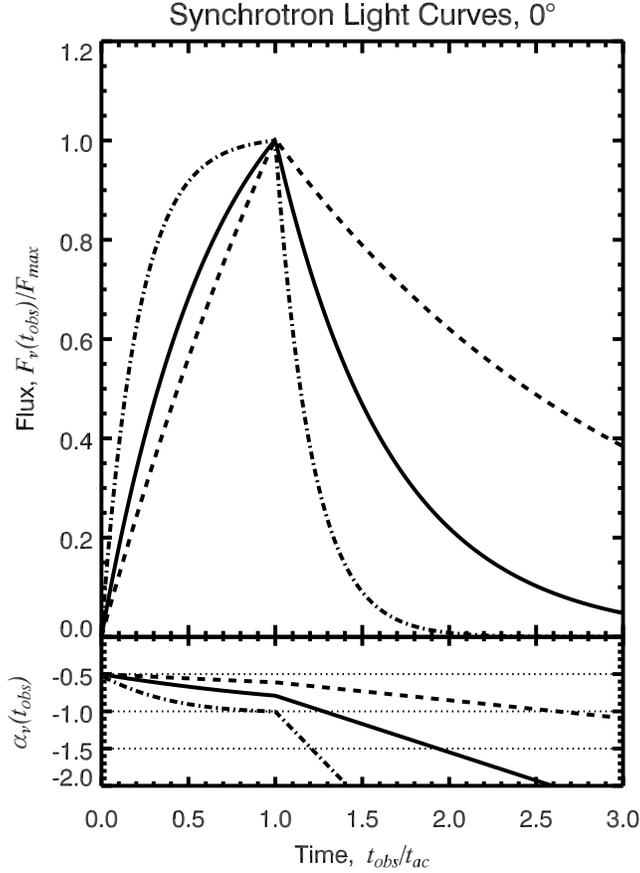}
\caption{Normalized synchrotron light curves for viewing angle $\theta_{obs}=0^{\circ}$
at frequencies of $10^{11}\,\mbox{Hz}$ (dashed), $10^{12}\,\mbox{Hz}$ (solid),
and $10^{13}\,\mbox{Hz}$ (dot-dashed)
in the rest frame of the shocked plasma.
The time dependence of the spectral index for each light curve is shown in the lower panel. 
The ratio of the synchrotron decay time and the apparent crossing time is 
$2.0$ ($10^{11}\,\mbox{Hz}$), $0.7$ ($10^{12}\,\mbox{Hz}$),
and $0.2$ ($10^{13}\,\mbox{Hz}$). 
The same input parameters were used as in Fig.~\ref{mssp.0}.
\label{mslc.0}}
\end{figure}

The spectral behavior is easier to interpret if one compares it with a set of complementary
light curves at a few representative frequencies.
At zero viewing angle (Fig.~\ref{mslc.0}), all light curves peak at the crossing time 
$t_{ac}=8\times10^6\,\mbox{s}$ for the parameters used (cf.~Fig.~\ref{mssp.0}).
However, only those light curves for which the decay time is close to the crossing time, $t_{\nu}\sim t_{ac}$,
will be roughly symmetric and have a sharp peak; i.e., one should expect symmetric flares only when 
the frequency of observation is close to the break frequency at $t_{obs}=t_{ac}$ for zero viewing angle.
At higher frequencies $\nu\gg\nu_b$,
one will observe a sharp rise followed by a flat top and then a quick decay.
At lower frequencies $\nu\ll\nu_b$, 
an increase in flux density
 that occurs on time scale $t_{ac}$ will be followed by a relatively long quasi-exponential decay
on time scale $t_{\nu}$.
In the latter case, one will observe the emission rise to a relatively stable high value of flux
with insignificant decay  if the period of observation is comparable to $t_{ac}$.
Although there is no time delay in the peak emission in the case of zero viewing angle,
a cross correlation analysis should detect
a positive time delay between high and low frequency synchrotron flares.
This is because  the flares at lower frequencies have larger decay times and do not
respond to changes in the source as quickly as the flares at higher frequencies;
i.e., lower frequency flares both rise and decay more slowly.

\begin{figure}
\plotone{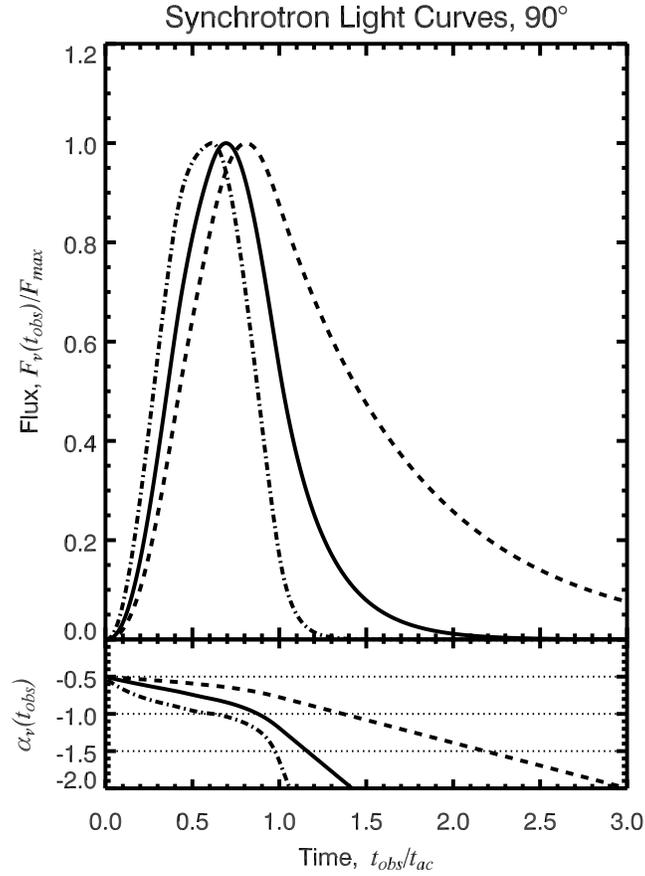}
\caption{The same as Fig.~\ref{mslc.0} but for viewing angle $\theta_{obs}=90^{\circ}$.
The light curve profiles are roughly symmetric over a broad range of frequencies.
The peak of the light curves at lower frequencies is delayed with respect to those at 
higher frequencies.
The ratio of the synchrotron decay time and the apparent crossing time is 
$0.8$ ($10^{11}\,\mbox{Hz}$), $0.3$ ($10^{12}\,\mbox{Hz}$),
and $0.08$ ($10^{13}\,\mbox{Hz}$). 
\label{mslc.90}}
\end{figure}

For viewing angle $\theta_{obs}=90^{\circ}$ (Fig.~\ref{mslc.90})
both the shapes and the positions of the peak emission are different
from the zero viewing angle case.
Due to the circular shape of the excitation region and the orientation of the
apparent shock front, the flares  appear symmetric even at frequencies higher 
than the break frequency
if the source is geometrically thin along the axis of the jet ($H\ll R$).
The peaks of the flares  occur between half the crossing time $t_{ac}/2$ and 
the crossing time $t_{ac}$, with higher frequency flare peaks tending toward $t_{ac}/2$ and
those at lower frequencies approaching $t_{ac}$.
The time delay of the peak emission is never expected to exceed half the crossing time
or roughly half the duration of the flare.
While comparing Figs.~\ref{mslc.0} and~\ref{mslc.90}
one should bear in mind that the crossing time $t_{ac}=2\times10^7\,\mbox{s}$
for viewing angle $\theta_{obs}=90^{\circ}$
is longer by 2.5 than that for zero viewing angle.

\subsection{Inverse Compton: Single emission region}
The inverse Compton (IC) emission coefficient in the SSC model results from the convolution of the
electron energy spectrum and the synchrotron emission spectrum.
The incident synchrotron intensity may depend strongly on the direction of the incident emission
$\vec\Omega_i$,
with each direction having its own set of characteristic frequencies $(\nu_t,\nu_b,\nu_d)$
for a specific time and location in the source.
However, the self-absorption frequency is not expected to depend as strongly on direction or time
and, therefore, its average value may be used in a qualitative analysis.
The turn-over frequency of the IC emission can therefore be estimated as
$\nu_{c,t}=4\gamma_{min}^2\langle\nu_t\rangle$, where $\langle\nu_t\rangle$ is the average 
self-absorption frequency of the internal synchrotron radiation field.
The synchrotron drop-off frequency $\nu_d$ for the directions terminated by the internal apparent shock front is 
constant and equal to the critical frequency of highest energy electrons, $\nu_{max}=c_1B\gamma_{max}^2$.
Thus, the IC emission is expected to drop off exponentially 
at frequencies greater than $\nu_{c,max}=4\gamma_{max}^2\nu_{max}$ as long as the internal shock front is
inside the excitation region.
Once the internal shock front exits, the IC drop-off frequency $\nu_{c,d}$
begins to decrease.
The break frequency of the incident synchrotron emission $\nu_b$ 
is 
a strong function of direction and time 
and thus 
does not translate into a well-defined break frequency of the IC spectrum.
Rather, we expect the IC spectrum to steepen 
gradually by $\Delta\alpha\approx1/2$
toward higher frequencies.

\begin{figure}
\plotone{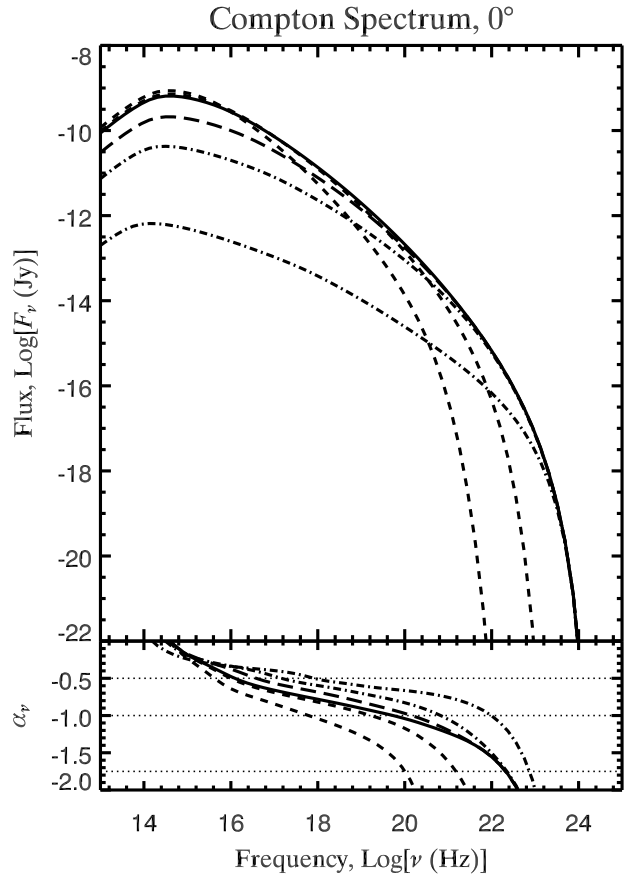}
\caption{The evolution of the inverse Compton spectrum for $\theta_{obs}=0^{\circ}$
in the rest frame of the shocked plasma in the forward shock region.
Compare with evolution of synchrotron spectrum in Fig.~\ref{mssp.0}.
\label{mcsp.0}}
\end{figure}

\begin{figure}
\plotone{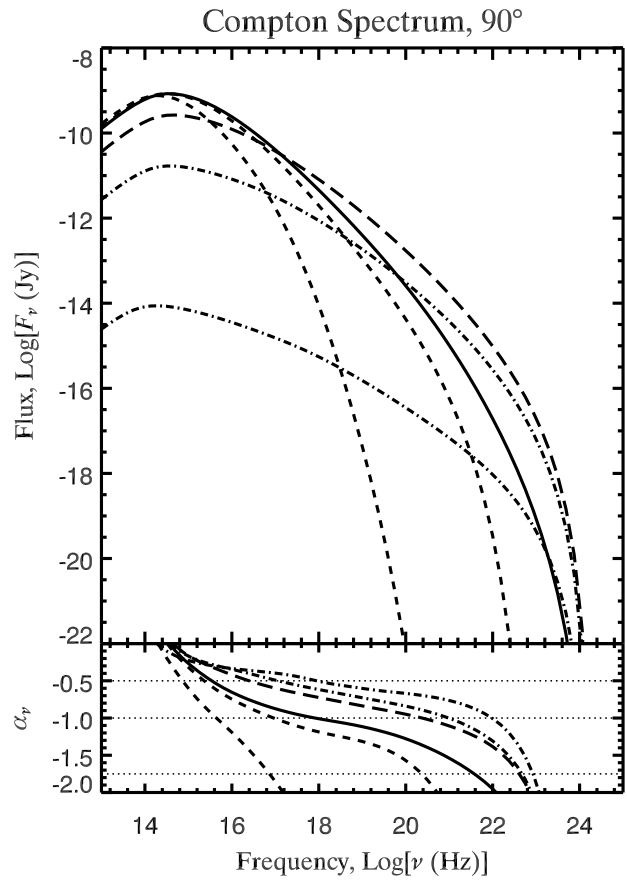}
\caption{The same as Fig.~\ref{mssp.0} but for $\theta_{obs}=90^{\circ}$.
Compare with  Fig.~\ref{mssp.90}.
\label{mcsp.90}}
\end{figure}

\begin{figure}
\plotone{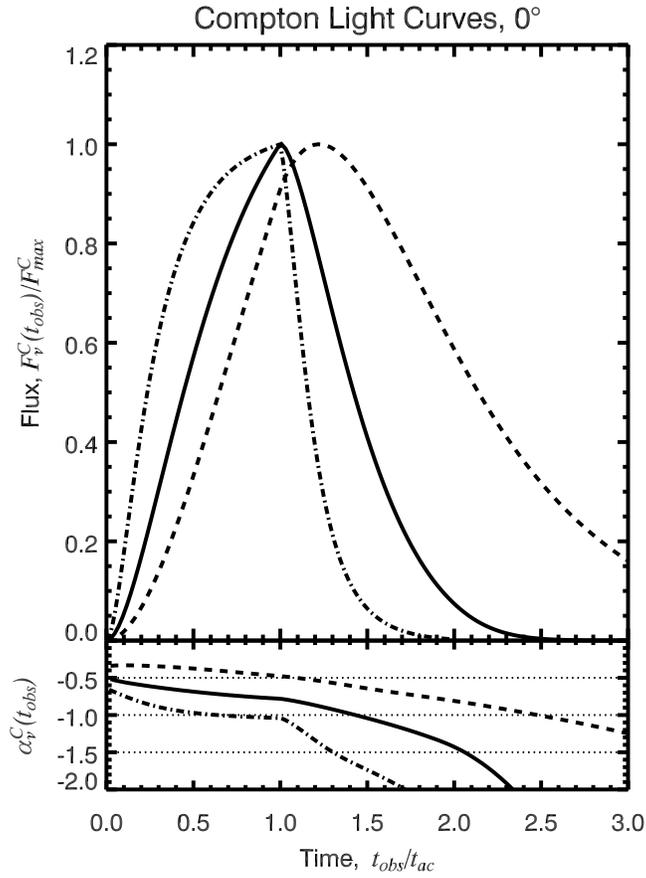}
\caption{Normalized inverse Compton light curves for viewing angle $\theta_{obs}=0^{\circ}$
at frequencies of $10^{16}\,\mbox{Hz}$ (dot-dashed), $10^{18}\,\mbox{Hz}$ (solid),
and $10^{20}\,\mbox{Hz}$ (dashed)
in the rest frame of the shocked plasma.
The time dependence of the spectral index for each light curve is shown in the lower panel. 
\label{mclc.0}}
\end{figure}

\begin{figure}
\plotone{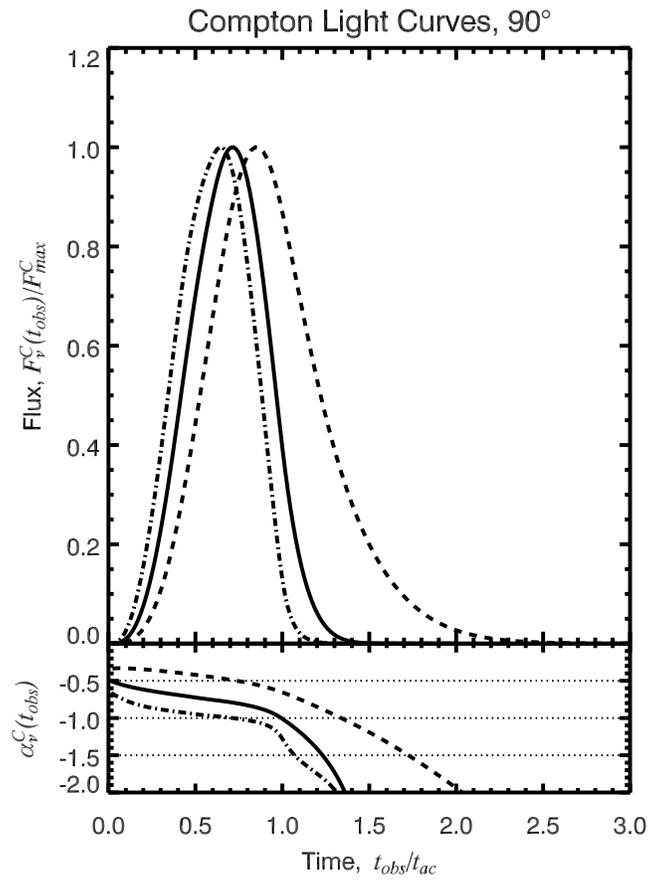}
\caption{The same as Fig.~\ref{mclc.0} but for viewing angle $\theta_{obs}=90^{\circ}$.
\label{mclc.90}}
\end{figure}

The evolution of the IC spectrum for two different viewing angles 
is shown in Figs.~\ref{mcsp.0} and~\ref{mcsp.90}. 
The overall behavior is similar to that of the synchrotron emission.
However, instead of the well-defined power-law seen in the synchrotron spectrum
the IC spectrum gradually steepens over a wide range of frequencies.
Hence, it cannot be approximated as a single  power-law
over more than two decades in frequency.
It is also clear that the IC emission peaks 
after the crossing time at lower frequencies (see dashed line in Fig.~\ref{mcsp.0}).
This pattern is more obvious if one examines a set of IC light curves at different frequencies,
as shown in Figs.~\ref{mclc.0} and~\ref{mclc.90}.
For viewing angle $\theta_{obs}=90^{\circ}$ (Fig.~\ref{mclc.90}), the behavior of the IC light curves is
very similar to that of the synchrotron flares.
The emission at higher frequencies tends to peak at time $\sim{}t_{ac}/2$.
The peak shifts toward the crossing time $t_{ac}$ at lower frequencies.
The light curves appear symmetric (similar rise and decay time) over a broad range of
frequencies.
This suggests that the internal light travel effects are insignificant for
this viewing angle and that the shapes of the flares are determined by
the geometry of the excitation region along the path of the apparent shock front.
However, in the case of zero viewing angle (Fig.~\ref{mclc.0})
the results of our simulations demonstrate that the IC flares
at sufficiently low frequencies peak after the crossing time,
in contrast to the synchrotron flares at this viewing angle, which all peak at
the crossing time.
Due to internal light travel delays, the IC flares can still increase in brightness after
the apparent shock front has left the excitation region and the high energy
electrons are no longer supplied.

The decay time of the IC emission is difficult to define uniquely because
it depends on the frequency-integrated incident synchrotron radiation, whose characteristics may depend  strongly
on direction, time, and location in the source.
Even when the position of the shock front as seen by a distant observer appears to have  left the source,
the internal shock front as viewed from a given location may still lie well inside the excitation region
for a large range of directions, thereby supplying synchrotron seed photons at the highest frequencies.
An upper estimate of the IC decay time can be obtained by assuming that
the maximum incident synchrotron frequency $\nu_{max}$ does not depend on time.
Then for a given IC frequency $\nu_c$ the decay time is the time needed for 
the drop-off frequency of the inverse Compton emission $\nu_{c,d}=4\gamma(\Delta{t})^2\nu_{max}$ 
to decrease to $\nu_c$. This gives the estimate 
\begin{equation}
t_{\nu_c}=t_1\sqrt{\frac{4\nu_{max}}{\nu_c}}.
\end{equation}
The actual decay time may be shorter because of the decrease in the drop-off frequency
of the incident synchrotron radiation.

\subsection{Time delays}
Comparison of the synchrotron and inverse Compton light curves offers a possible explanation 
of the existence of negative time delays between radio and X-ray flares.
In rough terms,  the time delays depend on (1) the apparent crossing time $t_{ac}$, which
is determined by the source dimensions and the viewing angle, and (2) the decay times
for synchrotron and IC emission.
If both decay times are much smaller than the crossing time, then no significant shifts
between these light curves will be  present.
Decreasing the frequency of observation of the synchrotron emission will lead to an increase
in the decay time.
If it becomes comparable to or longer than the crossing time,
the synchrotron flare will be delayed with respect to the IC flare.
If, on the other hand, one compares 
synchrotron emission at sufficiently high frequencies (infrared or optical) 
with
IC emission at relatively low frequencies (soft X-rays)
so that the corresponding IC decay time is longer than the crossing time,
then 
the IC flare should  peak after the synchrotron flare.

\subsection{Effects of the dual emission regions}
As a result of the collision between shocks both reverse and forward regions will be created.
Thus, the observer may see a combination of emission from these two regions.
For the most energetic collisions both shock  speeds should be close to $c/3$
in the rest frame of shocked plasma.
Within the assumed cylindrical geometry, one naturally expects the cylinder radius
to be the same for both regions, $R=R_r=R_f$, but the heights $H_r$ and $H_f$ may be different.

Due to intrinsic geometrical symmetries in the problem, one can conclude that
for viewing angle $\theta_{obs}=90^{\circ}$
the synchrotron emission observed from both  regions will be identical if 
all the input parameters of both regions are the same, i.e., if $H_r=H_f$, $n_r=n_f$, $B_r=B_f$, etc.
For all other viewing angles, 
if the apparent crossing times for both regions are equal, $t_{ac,r}=t_{ac,f}$, 
which imposes a constraint on the values of $H_r$ and $H_f$, and if
the other input parameters of both regions are identical (i.e., $n_r=n_f$, $B_r=B_f$, etc.),
then the optically thin emission from both regions will have identical shapes, albeit the flux level
should be different because of the difference between $H_r$ and $H_f$.
By adjusting $N_0=(s-1)n\gamma_{min}^{s-1}/[1-(\gamma_{min}/\gamma_{max})^{s-1}]$ 
one can  change the peak flux levels without
altering the shape of the light curve.
On the other hand, even moderate discrepancies in $s$, $\gamma_{max}$, or $B$ may lead to
dramatic differences in the profiles of the reverse and forward flares.

If observed flares in blazars stem from collisions between shocks,  then
all the flares must consist of two components (reverse and forward).
In most cases one of the components will overwhelm its counterpart,
resulting in an apparently single profile.
Furthermore, since blazar flares occur frequently, 
it is challenging to separate doubly peaked flares resulting from a single collision 
and two separate events occurring at about the same time.

\begin{figure}
\plotone{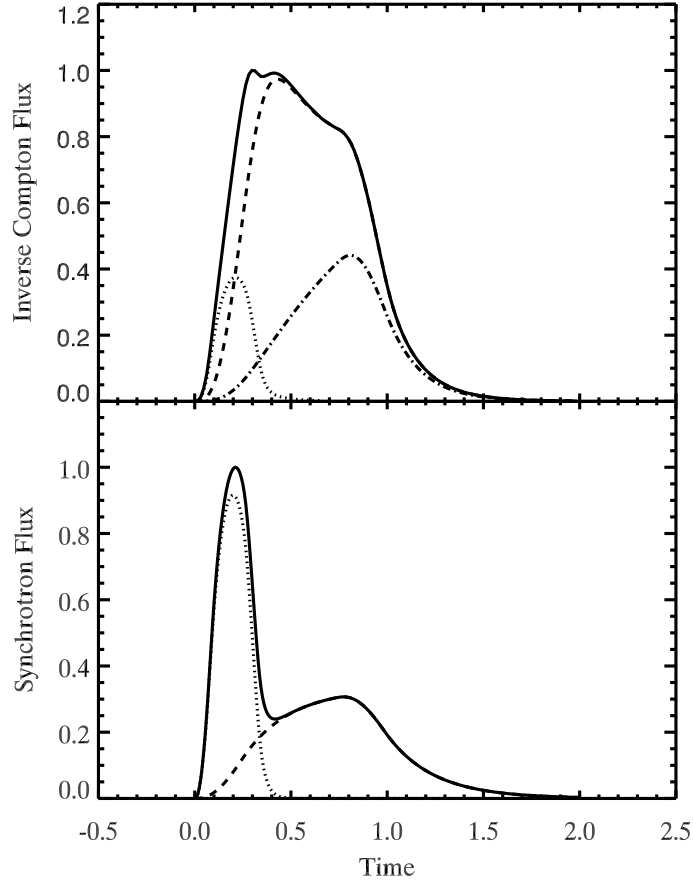}
\caption{Forward (dotted) and reverse (dashed) synchrotron ($10^{12}\,\mbox{Hz}$) 
and inverse Compton ($10^{17}\,\mbox{Hz}$) light curves
for viewing angle $\theta_{obs}=10^{\circ}$ in the plasma rest frame.
The time is normalized by  $t_{ac,r}=3\,t_{ac,f}$.
Solid curve is sum  of both forward and reverse light curves.
Dot-dashed curve corresponds to SSC emission from reverse region
when the contribution of the seed photons from the forward region is
neglected. 
We used the following parameters for the reverse region: $H=0.25\,R$, $B=1\,\mbox{G}$, $n=10\,\mbox{cm}^{-3}$;
for the forward region: $H=0.05\,R$, $B=4\,\mbox{G}$, $n=40\,\mbox{cm}^{-3}$.
The other input parameters are the same  for both regions:
$v=0.34c$, $\gamma_{min}=10$, $\gamma_{max}=2\times10^4$, $s=2$, $D_0=10^3\,\mbox{Mpc}$.
\label{msclcb}}
\end{figure}

To illustrate the possible morphology of a doubly peaked flare, we have conducted a simulation
with the forward region being denser but thinner than the reverse region (Fig.~\ref{msclcb}).
Synchrotron seed photons originating from the adjacent region were taken into account
in calculation of the  IC light curves.
Examination of the results indicates that
the seed photons in the reverse region are dominated by synchrotron emission 
from the adjacent forward region
because of the higher density and magnetic field in the latter.
The light travel delays for the seed photons from the adjacent region 
are also longer on average.
The choice of the input parameters results in a quick spike of synchrotron emission (forward region)
followed by a weaker but longer outburst (reverse flare).
The corresponding IC variability has a more complex profile
because the IC emission from the reverse region
is strongly affected by seed photons from the adjacent forward region. 
Notice that both flares begin at the same time, in accordance with our model.
The shock collision model cannot explain
a doubly peaked flare consisting of a broad weak outburst followed by a spike of emission at the end
because that would require different onset times for the reverse and forward flares.

\subsection{Application: 3C~273}
\begin{figure}
\plotone{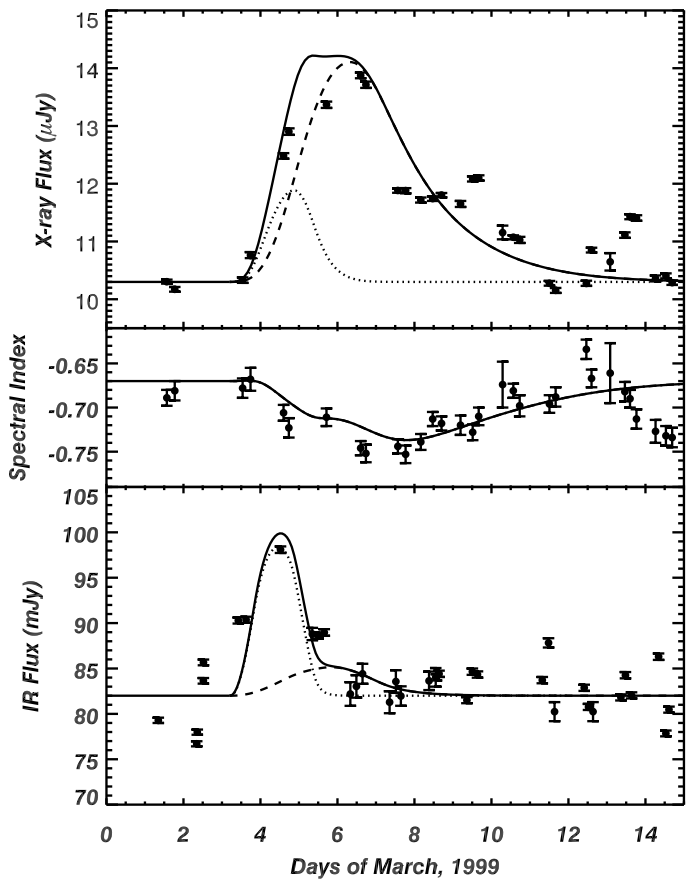}
\caption{\label{3C273.1999March}
A model fit to the March 1999 flare in 3C~273; 
the observed flux density at $2\,\mbox{keV}$ and $2.2\,\mu\mbox{m}$ is indicated with filled circles.
The observed and simulated X-ray spectral index variability is shown in the middle panel.
Contributions from the forward (dotted line) and reverse (dashed line) regions are shown.
A constant quiescent synchrotron flux of $F^S_{\nu}=82\,\mbox{mJy}$
is used based on the average K-band flux detected before and after the flare.
For the X-ray emission we use $F^C_{\nu}=10.3\,\mu\mbox{Jy}$
and spectral index $\alpha^C_{\nu}=0.67$ to describe a constant baseline during the flare.
We used the following input parameters: 
$H_r=0.6\times10^6\,\mbox{sec}$, 
$B_r=0.7\,\mbox{G}$,
$n_r=2.9\times10^3\,\mbox{cm}^{-3}$,
$\gamma_{max,r}=2\times10^3$ for the reverse region, and
$H_f=0.15\times10^6\,\mbox{sec}$, 
$B_f=2.0\,\mbox{G}$,
$n_f=4\times10^3\,\mbox{cm}^{-3}$,
$\gamma_{max,f}=4\times10^3$ for the forward region.
The other input parameters were identical for both regions:
$R=0.6\times10^6\,\mbox{sec}$,
$\gamma_{min}=50$, and $s=2.3$.
Modern cosmological parameters were used to convert the results of the simulation
into the observer frame: $H_0=70\,\mbox{km s}^{-1}/\mbox{Mpc}$, $\Omega_{\Lambda}=0.7$.}
\end{figure}

In March 1999 the quasar 3C~273 produced a short outburst that was observed simultaneously
in the infrared K-band and the X-ray band.
An examination of the  multifrequency data (Fig.~\ref{3C273.1999March}) 
shows that  the X-ray flare was delayed with respect
to its K-band counterpart by about one day. 
The activity in K-band spanned about two days, whereas the X-ray flare continued for about four days.
The spectrum of the X-ray emission was observed to steepen during the flare.

This long time delay can be explained by applying the shock collision scenario,
according to which the collision 
leads to the formation
of two adjacent emission zones with similar properties (reverse and forward regions).
Two conditions are required to produce a successful fit: 
(1) the X-ray flare is dominated by IC emission from a larger region
that has a crossing time of about four days,
and (2) the infrared flare is in turn dominated by synchrotron emission from a smaller  region
with apparent crossing time of about 2 days.
In general, these conditions are difficult to satisfy simultaneously since brighter synchrotron
emission will correspond to brighter IC emission.
However, if the Lorentz factor of the maximum energy electrons is limited to
$\gamma_{max}\sim10^3$, the magnetic field strength $B\sim1\,\mbox{G}$, 
and the bulk Lorentz factor of the emitting plasma $\Gamma\sim10$, 
then even a small discrepancy in $B$ and/or $\gamma_{max}$ 
between the forward and reverse regions may significantly affect the K-band synchrotron emission.

For simplicity and in order to minimize the number of freely adjustable parameters,
we adopt a  bulk Lorentz factor $\Gamma=10$
and angle between the jet axis and line of sight $\theta_{obs}=90^{\circ}$ in the plasma rest frame.
This viewing angle corresponds to $\theta^{\star}_{obs}=5.7^{\circ}$ in the observer's frame. 
We also assume that the speeds of both the forward and reverse shocks equal to $0.34c$ in the 
rest frame of the emitting plasma.
The observed duration of the flares combined with 
the direction of motion and the bulk speed of the plasma sets the size of the emission region at
about $0.5\times10^6\,\mbox{sec}$, whereas the roughly symmetric shape of the flares sets the magnetic field at
about $1\,\mbox{G}$. 
Adjusting these quantities and comparing the results of the simulation with the data 
culminate in the fit presented in Fig.~\ref{3C273.1999March}.

\section{Discussion}
\subsection{The Nature of Radio Core}
The model that we have developed can be seen as part of a more general picture of the jet 
in which a stationary system of oblique shock waves 
terminated by a Mach disk 
corresponds to the stationary feature called the radio core, 
which is seen in essentially all jets.
The model therefore relates the gas dynamical structure of the jet to its variability properties.
Provided that the shock collision scenario is correct, 
high speeds of superluminal components originating from the radio core
(bulk Lorentz factors of $10$ and higher) 
would  require relativistic shocks upstream of the stationary shock complex with bulk Lorentz factors
up to $200$ or even higher, since the collision will tend to substantially decelerate the moving shock.
This requirement of high upstream speed may be relaxed if interaction with the oblique shocks alone can
energize the moving shock sufficiently enough to explain the superluminal features,
since oblique shocks are not expected to slow down the moving shock appreciably and,
therefore, extremely high speeds prior to collision are not necessary.
We have only conducted a 1D study study of the collision, but the system of oblique shocks and 
Mach disk is inherently a 3D structure.
Therefore, more realistic simulations are required to ascertain the plausibility 
of our scenario and its gas dynamical consequences for relativistic jets.

The results of the modeling reported in this paper do not depend on
the details of how the excited plasma is produced. Most calculations are
performed in the rest frame of the emitting plasma, which gives them
considerable generality since one can transform to a frame in which the
plasma is moving.
Only the fit to the 3C~273 flare has been converted into the observer
frame to facilitate easy comparison with the observed light curves.
On the other hand, we have applied the model to a scenario in which the flares are
caused by a collision between a moving shock generated at the base
of the jet and a stationary Mach disk located downstream in the jet
at a distance of at least $\sim1\,\mbox{pc}$.
This collision scenario relies on the existence
of a stationary shock wave complex, and hence is different from
the ``internal shock'' scenario investigated by \citet{spa01},
who considered collisions between moving shells produced at the base of the jet.
The moving shock in our scenario and the shells in the model of
\citet{spa01} could both be produced at the base of the jet by similar
mechanisms, perhaps involving an instability in the accretion disk.
A shell of relativistic plasma is formed behind the moving shock
as it propagates downstream in the jet.
The collision of the moving shell with the Mach disk will occur
at roughly the same location in the jet, i.e., 
the location of the Mach disk, regardless of the frequency of ejections at the base of the jet. 
In contrast, a collision of two moving shells will occur
at different locations in the jet, since the time required for the
faster shell to catch up with the slower shell depends on the relative speeds and initial
separation of the shells. However, it can be easily shown that the ejection of two shells
separated by $\sim5\,\mbox{days}$ will result in the collision at the distance of $\sim1\,\mbox{pc}$ 
if the Lorentz factor of the slower shell is $\sim10$. 
Because of this, it might be difficult to distinguish
observationally between these two scenarios. Furthermore, our formalism can be applied
to the \citet{spa01} scenario with a suitable Lorentz transformation.

The bright appearance of the superluminal knots seen in VLBI images is at least partly
due to the relativistic speed of the plasma producing the emission. 
In the majority of blazars the radio core usually appears brighter than the superluminal features.
However, a stationary shock wave normal to the flow
of plasma in the jet slows the  flow down to a mildly relativistic speed.
Therefore, minimal relativistic boosting is expected in the case of quiescent emission from 
the plasma on the downstream side of the Mach disk.
The radio core is likely to be dominated by relativistic flow
passing through the system of oblique shocks or 
the reaccelerated flow farther downstream of the Mach disk.

The outcome of the collision depends on the relative speeds of the moving shock and
the quiescent flow between the shock and the Mach disk prior to collision.
If the speed of the moving shock prior to collision  is much higher
than the speed of the quiescent flow,
then the moving shock will destroy the stationary shock complex, 
resulting in an excited plasma propagating downstream at relativistic speed.
This scenario requires that the oblique shocks and the Mach disk reestablish
at approximately the same locations in the jet after the collision.
The time for this system to reappear in the jet is quite short \citep[see][]{gom97}.
The collision will manifest itself as an increase in brightness of the radio core
followed by the appearance of a superluminal knot associated with the shocked plasma
that drags the core downstream.
This must be followed by the reappearance of a bright feature at the original position of the radio core,
signifying the reestablishment of the Mach disk.
It might be possible to detect the temporary downstream translation of the apparent core
in high resolution, multi-epoch 
VLBI observations with phase-referencing to register the coordinates of the images. 
Alternatively, in the case of a weak moving shock, 
the stationary shock complex will be weakly affected by the collision.
It will retain its  position in the jet with probably only a slight shift downstream.
In this case, a superluminal knot associated with the flare can be identified with  the portion of
the moving shock that only interacts with oblique shocks 
between the Mach disk and the outer boundaries of the jet.

The existence and properties of the Mach disk established in 2D simulations and analytical
studies must be confirmed with high resolution 3D simulations.
We have assumed in our studies that the oblique shocks play an insignificant role in the observed variability.
However, existing 3D simulations indicate that strong oblique shocks can appear and
that the Mach disk may be small or even non-existent if
fluctuations of the jet direction are introduced \citep{alo03}.
This and other issues such as the time scale on which the stationary structure can be reestablished
should be addressed by future jet simulations.

\subsection{Emission calculation}
Our simulations  rely on approximations to the synchrotron 
emission and absorption coefficients that offer a number of advantages.
They allow a  more immediate yet comprehensive description of the excitation
properties of the medium energized by a propagating shock wave.
We have used these approximations to derive analytical solutions to the time-dependent
radiative transfer equations that describe the propagation of synchrotron emission
through an evolving relativistic medium with all light travel time effects taken
into account.
This significantly speeds up the IC  emission calculation.
For the cases involving simple geometry, such as a cylindrical source at zero viewing angle,
the equations can be integrated analytically.
Despite these advantages, the approximations can give erroneous results
at the later stages of the flare when the maximum electron energy $\gamma_{max}(\Delta{t})\rightarrow{}\gamma_{min}$.
To simplify the  analytical solutions, we have substituted $\exp(-[\Delta{t}/t_{\nu}]^2)$
with $\exp(-\Delta{t}/t_{\nu})$ 
to make the IC emission calculations with light travel delays for the seed photons possible. 
This procedure can underestimate the rate of decay of the observed synchrotron emission,
especially at low frequencies.
A more precise formulation can be worked out based on direct integration of $\exp(-[\Delta{t}/t_{\nu}]^2)$
and expression of the results  in term of  the Error function.

We have assumed that the electron energy losses are dominated by synchrotron emission.
Although this may not be the case in some sources, incorporation of inverse Compton losses
in a calculation that relies on a dynamical evaluation of the time-delayed 
synchrotron emission  poses considerable difficulties.
Indeed, if inverse Compton losses dominate then 
the electron distribution function at a given location and time in the source
will depend on the synchrotron radiation field at this location.
This photon field in turn depends on 
the structure of the electron energy distribution
throughout the source at retarded times, as  needed  to account for the  light travel delays.
Thus, the problem of determining the distribution of electrons 
not only becomes non-linear
but also non-local,
since the distribution of electrons at a given position will depend on the electron 
excitation structure in the whole emission region at different retarded times.
This task lies beyond the scope of the current work.

Source expansion, adiabatic energy losses, and gradients in magnetic field and
electron density have been excluded from the current work as well.
Consideration of these effects must be postponed until a more realistic 3D picture of
the collision becomes available.
The IC emission calculation relies on a number of analytical estimators
that apply to  a non-expanding source with uniform magnetic field and
density, a shock wave moving at constant speed, and evolving electron energy distribution
dominated by synchrotron energy losses.
Many of these estimators must be reconsidered if the source is allowed to expand 
substantially during the flare, or if gradients in magnetic field or electron density are introduced.

In this paper we have concentrated on the SSC mechanism for production of X-rays.
External Radiation Compton (ERC) emission, which is not affected by the light travel time effects
in the same manner as in the SSC model, will be considered in a separate work.
Since external emission in most cases is considered constant, it is expected to fill the source
uniformly.
Despite the absence of internal light travel delays in the ERC model, 
one cannot reject this model based solely on the presence 
of the relative time delays between multifrequency flares:
the internal excitation structure due to electron energy losses
can produce both positive and negative time delays between
synchrotron and IC variations for certain viewing angles,
most notably $\theta_{obs}\sim90^{\circ}$ in the plasma rest frame.
However, for viewing angles close to $\theta_{obs}=0^{\circ}$,
there is a fundamental distinction between SSC and ERC models.
The ERC emission is expected to peak at crossing time $t_{ac}$
similarly to the synchrotron flares at this viewing angle.
The SSC emission, on the other hand, can continue to brighten even after 
the crossing time because of the light travel delays of the internal synchrotron emission.
Mirror Compton emission, even if strong enough to compete with SSC emission,
is expected to be delayed by at  least several times $t_{ac}$
and is, therefore, completely disconnected from synchrotron and SSC variability.

A preliminary investigation indicates that the 3C~273 flare discussed above
cannot be explained by the ERC model with emission from the dusty torus as a dominant contribution.
The associated decay times tend to be higher than in the SSC model because
the maximum frequency of available seed photons is higher and does not decline with time,
resulting in flares with very long decay time, in contrast with observations.
The inclusion of adiabatic losses will decrease the time scale 
of variability of the ERC flare.
More importantly, the change in Doppler boosting of the external radiation
as the emission region propagates farther along the jet may quench
the ERC flare before any substantial decay of the energy of electron 
producing the emission.
These two effects must be significant for the flares of $\sim1$ month
or longer in duration.
However, their effect on the fast flares such as the one in 3C~273 
that we have considered may be neglected.

\subsection{Comparison with Observations}
Light travel delays strongly influence the multifrequency timing of
variable emission produced in relativistic jets of blazars.
Despite considerable complications that the light travel delays present in
modeling emission variability, they offer a number of diagnostics of
the emission regions that can be used in interpreting observational data.
In this section, we summarize some of these diagnostics.

Variability of synchrotron radiation at high frequencies 
originates from a thin emission zone around the shock front
propagating through the excitation region.
Thus, this emission traces the geometrical shape of the excitation region
along the path of the apparent shock front, whose orientation as a 
function of viewing angle is given by Eq.~(\ref{ttobs}).
For instance, for a viewing angle of $90^\circ$ in the plasma rest frame,
which  may correspond to only a few degrees in the observer's frame
if the jet is highly relativistic,
the apparent shock front is tilted with respect to the orientation
of the physical shock front (see Figs.~\ref{shock} and~\ref{apshock}).
Because of this tilt, the emission does not simply trace the geometry along the jet, 
but instead depends on the shape of
the excitation region both along and across the jet.
If the excitation region is thin ($H\ll R$), 
the shape of the observed flare will reflect its structure across the jet.
The geometry along the axis of the jet can only be probed in isolation
if the viewing angle is very small in the plasma rest frame, 
which requires almost perfect alignment of the jet axis and the line
of sight in the observer's frame.

\begin{figure}
\epsscale{0.8}
\plotone{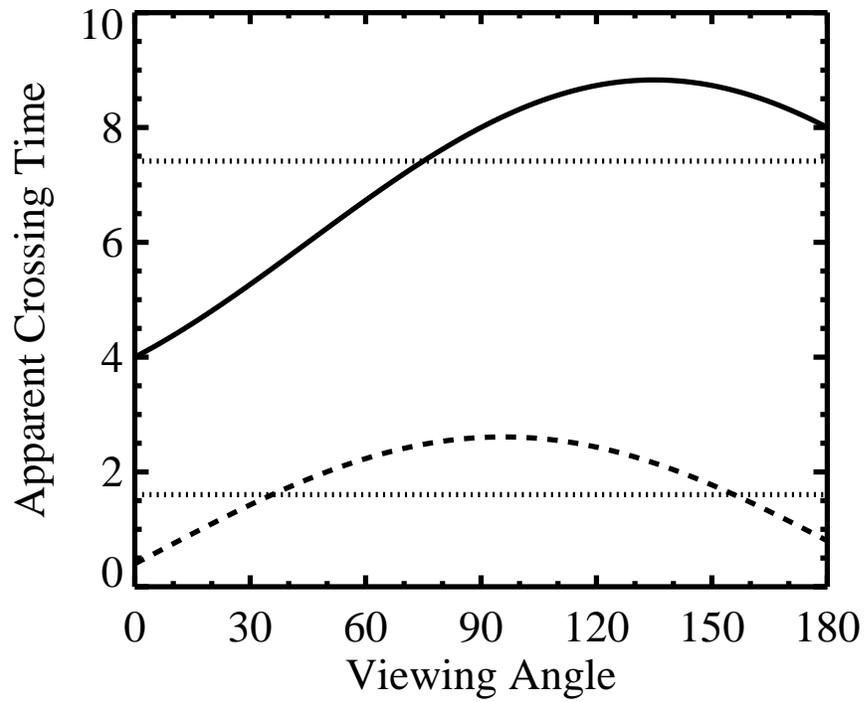}
\caption{Apparent crossing time $t_{ac}$ for $H=2R$ (solid) and $H=0.2R$ (dashed) 
as a function of viewing angle in the plasma rest frame.
We adopt $R=1$ and $v=1/3$.
Dotted lines at $t_{ac}=7.4$ and $t_{ac}=1.6$  correspond to 
the internal shock front crossing times $t_{int}$ for the central point (see text).
\label{delay}}
\end{figure}

Negative time delays between synchrotron and IC flares,
with variability at lower frequencies leading those at higher frequencies,
can be used to constrain the emission models.
As our simulations indicate, 
for relatively small viewing angles the internal light travel delays 
may exceed the apparent crossing time.
In this case, the IC flux at sufficiently low frequencies 
will continue to increase
for some period after the crossing time.
On the other hand, the ERC model is not expected to exhibit such behavior.
To estimate the range of viewing angles consistent with such behavior
we plot in Fig.~\ref{delay} the apparent crossing time $t_{ac}$ as a function
of viewing angle and an estimate of the internal light travel delays
based on the maximum time delay for the central point in the source $(0,0,H/2)$:
\begin{equation}
t_{int}=H/v+\left.\sqrt{R^2+(H/2)^2}\right/c.
\end{equation}
For a thin source ($H=0.2R$) the viewing angle in the observer frame must
be significantly smaller than $\sim35^\circ/(2\Gamma_{p}')$, where $\Gamma_{p}'$ is the bulk
speed of the emitting plasma.

For viewing angles $\theta_{obs}\sim90^\circ$, 
the internal light travel delays are relatively small
compared to the crossing time~$t_{ac}$.
For such viewing angles, frequency stratification and geometry
of the source will determine relative time delays between
flares at different frequencies.
In general, the ERC flares may exhibit similar behavior to 
the SSC flares since the electron excitation structure and
source geometry are the same.
However, the ERC model can still be rejected as  a viable X-ray
emission model for different reasons.
For example, if the external emission is provided by dust particles
in molecular torus at temperature $T_d=1000\,\mbox{K}$,
the seed emission will be dominated by infrared photons.
In this case, low energy relativistic electrons ($\gamma\sim15$)
are sufficient to produce X-ray emission in the RXTE band.
These electrons decay very slowly and, therefore,
the ERC flares observed with RXTE are expected to have
extremely long decay times, rendering them incapable
of fitting symmetric flares with roughly similar rise and
decay times as seen in 3C~273 (see Fig.~\ref{3C273.1999March}).

The internal light travel delays are emphasized even further 
when both reverse and forward zones are taken into account, 
since some seed photons should arrive from the adjacent region,
which will require additional time.
Because of the seed emission originating in the adjacent zone,
IC flares may possess complex profiles
that cannot be described as a sum of two independent flares
in the same way as can be done with the synchrotron flares.
Moreover, as our modeling of the 3C~273 flares indicates,
longer observed time delays can be explained with SSC emission
from dual regions than is allowed by single region
modeling.
In particular, it is possible to explain SSC flares
that peak not only after the maximum of the synchrotron emission,
but also when the synchrotron emission has virtually returned
to the background value that existed prior to the flare.
Further investigation of SSC flare profiles
from dual emission regions and their dependence on viewing angle will
be considered in a later paper.

\section{Summary}
We have developed a model of the emission regions in blazar jets responsible for
the fast outbursts of flux that
characterize this class of active galactic nuclei.
The collision of a moving shock wave
and a stationary shock complex in the relativistic jet 
results in a pair of propagating  shocks that drive acceleration of particles 
and strengthen the magnetic field,
thus producing enhanced synchrotron and IC emission.
The model incorporates time-dependent radiative transfer
through a rapidly evolving relativistic medium 
to find the time-delayed synchrotron radiation field 
that provides seed photons for IC scattering.
The ability of the model to trace the evolution of the excitation structure in the source
provides a powerful tool for probing the inner geometry of the jet structures
responsible for the sudden outbursts of activity usually associated with the ejection of
superluminal features from the radio core.
The model is especially well suited for modeling multifrequency variability
involving relative time delays between light curves in
different spectral bands.

The investigation of internal light travel delays that
apply to  the seed emission in the SSC model
strongly suggests that the relative time delays
between synchrotron and inverse Compton flares can be 
dominated by this effect.
The double structure of the excitation region resulting from
a shock collision can explain extreme reverse time delays
when the peak of the inverse Compton emission occurs after
the synchrotron flare, which starts at the same time,
has returned to the background level.

\acknowledgments
This research was supported through NASA grant NAG5-11811 and National Science Foundation
grant AST-0098579.

\appendix

\section{The Apparent Position of the Shock Front}
\label{appappshock}
\begin{figure}
\plotone{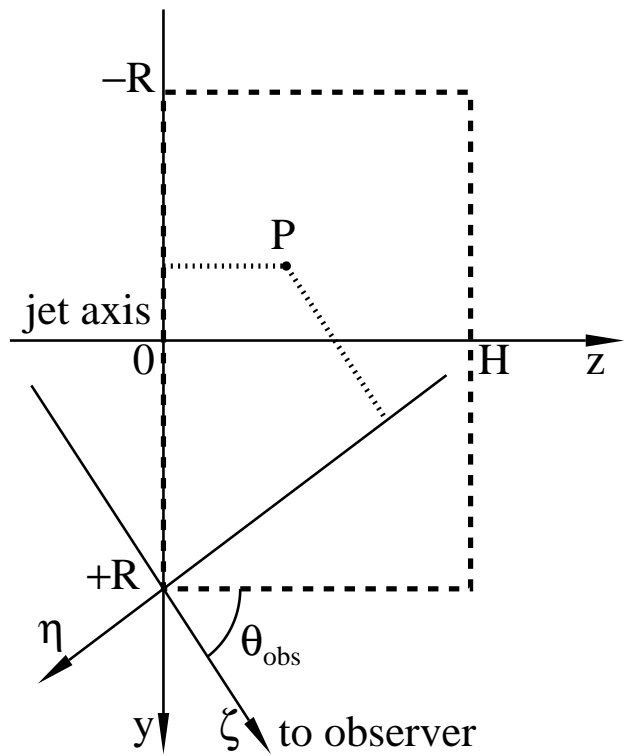}
\caption{Geometry for determining the apparent position of the shock front.\label{apfront}}
\end{figure}

Expression~(\ref{fapshock})  for the apparent position of the excitation front  may be derived as follows.
We define coordinate system $(\xi,\eta,\zeta)$  and 
consider emission from an arbitrary point $P=(x,y,z)$ within the boundaries of the source (see Fig.~\ref{apfront}).
The coordinate system $(\xi,\eta,\zeta)$ is obtained from the $(x,y,z)$ coordinates
by shifting the origin to $(0,R,0)$ and rotating the shifted $z$-axis by angle $\theta_{obs}$
in the $x=0$ plane such that the $\zeta$ axis points toward a distant observer in the plasma rest frame.

The shock front reaches point $P$ at time $t_1=z/v$ and excites
the electrons there.
A  time interval $t_2=[(R-y)\sin{\theta_{obs}}-z\cos{\theta_{obs}}]/c$ later, the radiation
produced at $P$ reaches the $\zeta=0$ plane that intersects point $(0,R,0)$ 
and is normal to the line of sight.
A further time interval $t_3=D_0/c$ is required for this radiation to reach the observer at
distance $D_0$ from the plane $\zeta=0$.
Thus, the observer will detect radiation from point $P$ at time
\begin{equation}
t_{obs}=t_1+t_2+t_3-D_0/c=z/v+[(R-y)\sin{\theta_{obs}}-z\cos{\theta_{obs}}]/c.
\end{equation}
The time $t_{obs}$ is defined such that the observer detects emission
from point $(0,R,0)$ at time $t_{obs}=0$.
Rearranging terms in this equation yields
\begin{equation}
z(1-v\cos{\theta_{obs}}/c)=(y-R)v\sin{\theta_{obs}}/c+v t_{obs},
\end{equation}
which defines the apparent shock front position at time $t_{obs}$.

\section{The Redistribution Function and Limits of Integration}
\label{appredist}
In general the limits of integration in Eq.~(\ref{compcoef}) are determined by conditions under
which it is energetically possible for an electron  with a Lorentz
factor $\gamma$ to scatter a photon of frequency $\nu_i$ moving in direction $\vec\Omega_i$ into
new direction $\vec\Omega_f$ with new frequency $\nu_f$.
For an isotropic distribution of electrons, the condition is
\begin{equation}\label{cond.ic}
(\gamma^2-1)
[(e_i-e_f)^2+2e_ie_f(1-\mu_{s})]
\ge
[\gamma(e_i-e_f)-e_ie_f(1-\mu_{s})]^2,
\end{equation}
where $e_{i,f}=h\nu_{i,f}/(m_ec^2)$ and $\mu_{s}=\vec\Omega_i\cdot\vec\Omega_f$.

We will consider the limits of the integration in Eq.~(\ref{compcoef}) in the Thomson limit
when $\gamma{}e_i\ll1$.
This results in pure Thomson scattering in the rest frame of
the scattering electron so that the photon frequency does not change in this frame of reference.
The relativistic transformations from the rest frame of the plasma to that of the scattering electron
and back to the plasma rest frame are the only factors responsible for the 
photon energy change in the Thomson limit.
The highest synchrotron photon frequency available for scattering in the SSC model
is  $\nu_{i,max}\sim{}c_1B\gamma_{max}^2$.
For $\gamma_{max}=10^4$ and $B=1\,\mbox{G}$ one finds $\nu_{i,max}\sim4.2\times10^{14}\,\mbox{Hz}$
and $\gamma_{max}e_{i,max}\approx0.02\ll1$.
Thus, for typical values of the magnetic field and maximum electron energy found in blazars,
inverse Compton scattering occurs in the Thomson limit.

Applying formula~(\ref{cond.ic}) to the particular integration order in Eq.~(\ref{compcoef}) 
provides the following limits.
Both outer integrals are taken over the complete range of possible values, i.e.,
$\vec\Omega_i$ is integrated over the entire sphere and $\gamma$ is integrated
from $\gamma_{min}(\Delta{t})$ to $\gamma_{max}({\Delta{t}})$.
In practice, the lower limit on $\gamma$ may depend on the maximum available synchrotron frequency
and the final frequency of scattering.
Given the final frequency $\nu_f$, scattering angle $\mu_{s}$, 
and scattering electron Lorentz factor $\gamma$,
the range of energetically possible values of the initial frequency $\nu_i$ is limited to
\begin{eqnarray}
&\displaystyle
\nu_f/A<\nu_i<A\nu_f,\\
\nonumber
&\displaystyle
\mbox{where}\quad A=[1+(\gamma^2-1)(1-\mu_{s})]+\sqrt{[1+(\gamma^2-1)(1-\mu_{s})]^2-1}.
\end{eqnarray}
For $\mu_{s}=1$, i.e., no change in direction of the photon, the
only allowed value is $\nu_i=\nu_f$; no scattering can effectively occur in this case.
On the other hand, for $\mu_{s}=-1$ the maximum possible range of the final frequencies is
achieved, namely $\nu_f/(4\gamma^2)\le\nu_i\le4\gamma^2\nu_f$ for $\gamma\gg1$.
In this case, the largest gain of the photon's energy $e_f=4\gamma^2 e_i$ 
occurs in a head-on collision with an electron,
while the photons that approach electrons from behind lose most of their energy:
$e_f=e_i/(4\gamma^2)$.
In practice, the upper limit on $\nu_i$ will be determined by $\nu_{max}$.

Applying relativistic transformations, combined with the Klein-Nishina cross-section,
to the problem of scattering results in a rather complicated integral expression
that includes the incident radiation spectrum and the full distribution function
of electrons, with arbitrary directional dependence.
The latter was integrated out by \citet{tra97} for isotropic electron distribution,
which led to the following expression for the redistribution function:
\begin{eqnarray}
&
\nonumber
\displaystyle
\psi
=
\frac{r_e^2}{2\nu_i\gamma^2}
\left\{
  \frac{1}{\sqrt{1-2\mu_s e_i/e_f+(e_i/e_f)^2}}
  -
  \frac{e_f(e_i+e_f)(2\gamma+e_i-e_f)}{2a_ia_f(a_i+a_f)}
  \times
\right.
  \\
&
\nonumber
\displaystyle
  \times
  \left[
    \left(
      \frac{2}{e_ie_f(1-\mu_s)}-1
    \right)
    -
    \frac{1}{(e_ie_f(1-\mu_s))^2}
    \left(
      (e_i+e_f)^2\gamma(2\gamma+e_i-e_f)\frac{(a_i+a_f)^2+a_ia_f}{(a_ia_f(a_i+a_f))^2}
      -
    \right.
  \right.
      \\
&
\displaystyle
\left.
  \left.
    \left.
      -
      \left(
        e_i^2\frac{2a_i+a_f}{a_i^2(a_i+a_f)}
	-
	e_ie_f\mu_s\frac{(a_i+a_f)^2-a_ia_f}{(a_ia_f)^2}
	+
	e_f^2\frac{a_i+2a_f}{a_f^2(a_i+a_f)}
      \right)
    \right)
  \right]
\right\}.
\end{eqnarray}
Here, $a_i=\sqrt{(\gamma+e_i)^2-(1+\mu_s)/(1-\mu_s)}$, $a_f=\sqrt{(\gamma-e_f)^2-(1+\mu_s)/(1-\mu_s)}$,
and $r_e$ is the classical electron radius.
In the Thomson limit $\gamma{}e_i\ll1$.
If one additionally assumes that $\nu_i\ll{}\nu_f$
and $\gamma\gg1$,
the redistribution function reduces to
\begin{equation}
\psi
=
\frac{r_e^2}{2\nu_i\gamma^2}
\left\{
1
-
\frac{\nu_f}{\gamma^2\nu_i(1-\mu_s)}
+
\frac{\nu_f^2}{2\gamma^4\nu_i^2(1-\mu_s)^2}
\right\},
\end{equation}
which is equivalent to the result obtained by \citet{rey82}.

\subsection{Evaluating the Inverse Compton Emission Coefficient}
The multi-integral expression for the inverse Compton emission coefficient is frequently
simplified using a $\delta$-function approximation of the incident emission
and/or the electron spectrum.
We will demonstrate here that consideration of the broad-band incident spectrum is necessary.

At a given location $(t,x,y,z)$ the spectrum of the incident synchrotron emission in direction $\vec\Omega_i$
is given by Eqs.~(\ref{insynch}) and~(\ref{insynch.tau}).
Analysis of this solution shows that it can be represented roughly as
\begin{equation}
I_{\nu_i}\propto
\left\{
\begin{array}{ll}
\nu^{5/2},&\nu<\nu_t,\\
\nu^{-(s-1)/2},&\nu_t<\nu<\nu_b,\\
\nu^{-s/2},&\nu_b<\nu<\nu_d,\\
\exp(-\nu/\nu_d),&\nu_d<\nu,
\end{array}
\right.
\end{equation}
where $\nu_t$, $\nu_b$, $\nu_d$ are the self-absorption frequency, the break frequency,
and the exponential drop-off frequency, respectively.
We have tacitly assumed that $\nu_t<\nu_b<\nu_d$, but one can readily visualize the shape of the resulting
spectrum if the relations were different.
The largest contribution to the incident synchrotron emission should be expected to
arrive from directions that intersect the internal apparent position of the shock front given by Eq.~(\ref{apshockf}).
The self-absorption frequency should increase only slightly 
due to an increase in the distance $L$ between the internal apparent shock front position
and the point $(x,y,z)$ at which the emission coefficient is evaluated, $\nu_t\propto L^{2/(s+4)}$.
The break frequency is expected to drop significantly, $\nu_b\propto L^{-2}$,
until it reaches its minimum value when the apparent location of the shock reaches the geometric boundary
of the excitation region.
On the other hand, the drop-off frequency should remain constant and equal to $\nu_{max}=c_1B\gamma_{max}^2$
as long as the apparent shock is inside the excitation region.
Once the shock reaches the boundary the decrease in $\nu_d$ will follow the changes in $\gamma_{max}(\Delta{}t)$,
where $\Delta{}t$ is the time since the apparent shock crossed the geometric boundary.

Although the redistribution function $\psi$ has rather complex dependence on 
electron Lorentz factor and the incident and final frequencies, one can very crudely approximate
it as $\psi\sim r_e^2/(2\nu_i\gamma^2)$ in the Thomson limit.
The innermost integral in the expression for the inverse Compton emission coefficient
can now be estimated by substituting these approximations and performing the integration
over $\nu_i$ from $\nu_f/(2(1-\mu_{s})\gamma^2)$ to $\nu_f$. This  gives
\begin{equation}\label{2ndint}
N(\gamma)\int{}d\,\nu_i{}\frac{h\nu_f}{h\nu_i}I_{\nu_i}\psi\propto
\left\{
\begin{array}{ll}
\exp(-(\gamma_d/\gamma)^2),& \gamma<\gamma_d\\
\nu_f^{-s/2}\gamma^0,& \gamma_d<\gamma<\gamma_b\\
\nu_f^{-(s-1)/2}\gamma^{-1},& \gamma_b<\gamma<\gamma_t\\
\nu_f\gamma^{-(s+2)},&\gamma_t<\gamma,
\end{array}
\right.
\end{equation}
where the critical Lorentz factors $(\gamma_{d},\gamma_{b},\gamma_{t})$ are found by solving
\begin{equation}
\nu=\frac{\nu_f}{2(1-\mu_{s})\gamma^2}
\end{equation}
for $\gamma$ with $\nu=\nu_d$, $\nu_b,$ and $\nu_t$.
Function~(\ref{2ndint}) is then integrated over $\gamma$ from $\gamma_{min}(\Delta{t})$ 
to $\gamma_{max}(\Delta{t})$.
From this solution it is apparent that electrons with Lorentz factors $\gamma_d<\gamma<\gamma_b$ and
synchrotron seed photons with frequencies $\nu_b<\nu<\nu_d$ provide equal contributions
to the IC emission coefficient \citep[see also][]{mch99}.
This is due to the fact that 
more abundant lower energy photons must scatter off of higher energy electrons, which are rarer,
in order to contribute to the IC emission at  $\nu_f$.
Yet, less common higher energy seed photons can lead to the same final frequency $\nu_f$
after being scattered by more plentiful lower energy electrons.

\clearpage 

\clearpage 

\clearpage 

\clearpage

\clearpage

\clearpage

\clearpage 

\clearpage 

\clearpage 

\clearpage 

\clearpage 

\clearpage 

\clearpage 

\clearpage 

\clearpage 

\clearpage 

\clearpage 

\clearpage 

\end{document}